\documentclass[journal]{IEEEtran}

\usepackage[utf8]{inputenc}
\usepackage[T1]{fontenc}  
\usepackage{amsmath,amssymb,amsfonts}
\usepackage{algorithmic}
\usepackage{graphicx}
\usepackage{subfig}
\usepackage[acronym]{glossaries}
\usepackage{cite}
\usepackage{braket}
\usepackage{caption}
\usepackage{float}
\usepackage[hyphens]{url}
\usepackage{hyperref}

\newacronym{aom}{AOM}{Acousto-optic modulators}

\newacronym{bs}{BS}{beamsplitters}

\newacronym{cdc}{CDC}{clock domain crossing}
\newacronym{cv}{CV}{continuous variables}

\newacronym{dma}{DMA}{Direct Memory Access}
\newacronym{dv}{DV}{discrete variables}

\newacronym{eom}{EOM}{Electro-optic modulators}

\newacronym{ff}{FF}{feedforward system}
\newacronym{fess}{FESS}{Feedforward Enabled Solid State Simulator}
\newacronym{fifo}{FIFO}{First In First Out}
\newacronym{fpga}{FPGA}{Field Programmable Gate Arrays}
\newacronym{fdb}{FDB}{FPGA Driver Board}

\newacronym{gbs}{GBS}{Gaussian Boson Sampling}
\newacronym{gth}{GTH}{Gigabit Transceiver High-performance}

\newacronym{hqe}{HQE}{high quantum efficiency}
\newacronym{hd}{HD}{homodyne detector}

\newacronym{im}{IM}{intensity modulator}
\newacronym{iqhe}{IQHE}{integer quantum Hall effect}

\newacronym{lwip}{lwIP}{lightweight IP}
\newacronym{lo}{LO}{local oscillator}

\newacronym{mbgbs}{MB-GBS}{Measurement-Based Gaussian Boson Sampling}
\newacronym{mbqc}{MBQC}{Measurement-Based Quantum Computing}
\newacronym{mbqip}{MB-QIP}{measurement-based quantum information processing}
\newacronym{mmcm}{MMCM}{mixed-mode clock manager}

\newacronym{pll}{PLL}{phase-locked loop}
\newacronym{pm}{PM}{phase modulator}
\newacronym{ps}{PS}{phase shifters}
\newacronym{pl}{PL}{Programmable Logic}
\newacronym{ps_fpga}{PS}{Processing System}

\newacronym{sss}{SSS}{Solid State Simulator}
\newacronym{ssh}{SSH}{Su–Schrieffer–Heeger}

\newacronym{tmf}{TMF}{temporal mode function}

\newacronym{qe}{QE}{quantum efficiency}

\usepackage{textcomp}
\def\BibTeX{{\rm B\kern-.05em{\sc i\kern-.025em b}\kern-.08em
    T\kern-.1667em\lower.7ex\hbox{E}\kern-.125emX}}
\begin{document}

\title{FPGA Based
Feedforward System for Photonic Quantum Computing Applications}
\author{
Daniel Duggan$^{1,2}$,
Simon Filgis$^{1,3}$,
Axel B. Bregnsbo$^{2}$,
Jürgen Saalmüller$^{1}$,
Jonas S. Neergaard-Nielsen$^{2}$,
Tobias Wintermantel$^{1}$,
and Ulrik L. Andersen$^{2}$
\thanks{$^{1}$ Q.ANT GmbH, 70565 Stuttgart, Germany.}
\thanks{$^{2}$ Center for Macroscopic Quantum States (Big Q), Department of Physics, Technical University of Denmark, 2800 Kongens Lyngby, Denmark.}
\thanks{$^{3}$ Ingenieurbüro Filgis, D-81677 Munich, Germany.}
\thanks{Corresponding author: Daniel Duggan (daniel.duggan@qant.gmbh)}
}

\maketitle

\begin{abstract}

Field-programmable gate arrays provide a high-performance solution for real-time signal processing in emerging quantum and photonic technologies. We present an FPGA-based fast feedforward system, that incorporates a high quantum efficiency fully fibre based homodyne detector, to enable low-latency signal processing critical for \gls{cv} \gls{mbqip} protocols. \gls{cv} \gls{mbqip} typically relies on adaptive measurements and/or displacements via feedforward to achieve scalability and universality, but existing implementations typically handle these operations in post-processing, limiting real-time applicability. Our system performs signal acquisition, conditioning, and logic operations in real-time, meeting the tight latency requirements of photonic quantum computing protocols. The detector exhibits a large clearance of 15 dB at 1 GHz with 4 mW linear oscillator and quantum efficiencies of $>$ 95\% with a total system latency of 196 ns. This work highlights the role of FPGAs in bridging the gap between theoretical models and physical implementations in photonics-based technologies.
\end{abstract}

\section{Introduction}
\label{sec:introduction}
The science of photonics is a powerful tool for creating new technologies and deepening our understanding of the world using light. In particular, photonic simulation and computing allow researchers to model and investigate quantum phenomena with high precision, while offering a promising platform for building scalable, room-temperature quantum processors. In this context, quantum computing has emerged as a particularly prominent area of research and development.

While many architectures exist to realise a useful quantum computer, one particularly promising approach is cluster state computation\cite{Rauss_1wayQC}, also known as \gls{mbqc}, which has gained attention due to its inherent scalability\cite{larsen_det_gen}\cite{Asavanant}. In this model, a highly entangled optical state, termed a cluster state, is generated, measured, and manipulated using feedforwarding of the measurement results to implement algorithms. Cluster states can be constructed using either \gls{dv}\cite{dv_clus_state} or \gls{cv}\cite{larsen_det_gen} encodings, with the latter leveraging the quadratures of light fields to represent quantum information and results in the field of \gls{cv}-\gls{mbqc}. While feedforwarding can be done in post-processing, real-time feedforwarding is essential for executing quantum algorithms in a fault-tolerant and scalable way.

The required feedforwarding demands fast manipulation of detected signals with minimal latency and high timing precision. \gls{fpga}s are well-suited to this task due to their inherently parallel architecture, reconfigurability, and low-latency operation. Unlike traditional CPUs or GPUs, which process instructions sequentially, FPGAs can implement custom logic pipelines that process incoming signals and generate control outputs within nanoseconds. This makes them ideal for real-time quantum photonic applications, where measurements directly influence future operations on ultrafast timescales. 

Alas, the purpose of this paper is three-fold:

\begin{enumerate}
    \item to introduce \gls{cv} \gls{mbqip} to a non-physics audience, and specifically electrical engineers. Whilst an introduction to \gls{mbqip} aimed at this audience does exist\cite{Scott_2022_TQE}, this paper discussed \gls{mbqip} using \gls{dv}; not \gls{cv} as we do here. 
    \item to serve as a reference for physicists on the electronic requirements of feedforwarding, and for those without FPGA experience to understand its developmental needs.
    \item to present an \gls{fpga}-based fast \gls{ff} that enables real-time feedforwarding of measurement results for photonic computing.
\end{enumerate}   

The system offers a high degree of experimental control, including precise timing, programmable scaling, and real-time modification of measurement outcomes. Built around a Xilinx ZCU102 development board, the platform consists of a \gls{hqe} fully fibre-based \gls{hd}, and integrates high-speed analog-to-digital and digital-to-analog conversions, allowing for low-latency processing of photonic signals. Measurement outcomes are processed on the FPGA and conditionally fed forward to optical modulators within sub-microsecond timescales. This makes the system well-suited for both \gls{mbqip}, and for photonic simulations that require dynamic, real-time control of lattice evolution. 

The structure of this article is as follows. In section \ref{sec:s2} we give an overview of the required physical concepts required to understand \gls{cv} \gls{mbqip}. Section 3 details the \gls{ff} system itself including its architecture and functional verification. Section 4 details potential improvements and discussion before we conclude.

\section{Physical Background}\label{sec:s2}
\subsection{Measurement Based Quantum Information Processing}

We first discuss \gls{mbqip}. In this architecture, algorithms are implemented by making homodyne measurements on a resource state called a cluster state. This state can be created using a fixed number of optical components. This is in contrast with the 'gate' based model, which typically uses optical components and single photon detectors to carry out algorithms. The number of required optical components depends on the complexity of the algorithm. In \gls{cv}, information is carried by the quadratures which we introduce first. To understand the need for feedforwarding, we introduce Gaussian states and homodyne detection. 

\subsubsection{Quadratures}
\gls{mbqip} can be performed using \gls{dv} or \gls{cv}. The difference lies in how information is encoded. For \gls{dv}, information is encoded in a two-level system termed a qubit. This can be physically realised in a photonic degree of freedom that has a discrete set of eigenvalues such as polarization, time-bins, or paths (dual-rail qubits).  For \gls{cv}, information is encoded in qu-modes i.e. a physical system that represents a quantum harmonic oscillator. Here, the degree of freedom has a continuous set of eigenvalues, and this degree of freedom is nearly always the aforementioned light field quadratures. Experimentally, one main difference between \gls{dv} and \gls{cv} manifests in how this information is measured: \gls{dv} implementations rely on single-photon detectors to project onto discrete basis states, while \gls{cv} implementations use homodyne detection to measure continuous quadrature values of the electromagnetic field.

To introduce the quadratures, we follow the pedagogy of \cite{Gerry_Knight_IntroQO} and assume a standing wave radiation field confined inside a 1-D cavity along the z-axis with perfectly reflecting mirrors. Assuming polarisation along the x-axis we have 

\begin{equation}
        \mathbf{E}_{\mathbf{r},t} = \mathbf{e}_{x}E_{x}(z,t).
\end{equation}

Solving Maxwell's equations without sources gives a single-mode field

\begin{equation}\label{eqt:e_field}
    E_{x}(z,t) = (\frac{2\omega^{2}}{V\epsilon_{0}})^\frac{1}{2}q(t)\sin(kz)
\end{equation}

\noindent where $\epsilon_{0}$ is the permittivity of free space, $\omega$ is the frequency of the mode, $k = \frac{\omega}{c}$ is the wave number and $q(t)$ acts as a canonical position. The cavity magnetic field is

\begin{equation}\label{eqt:cla_mag}
    \mathbf{B}_{\mathbf{r},t} = \mathbf{e}_{y}\frac{\mu_{0}\epsilon_{0}}{k}(\frac{2\omega^{2}}{V\epsilon_{0}})^{\frac{1}{2}}\rho(t)\cos(kz) = \mathbf{e}_{y}B_{y}(z,t) 
\end{equation}

\begin{figure}[t]
  \centering
  \includegraphics[width=\linewidth]{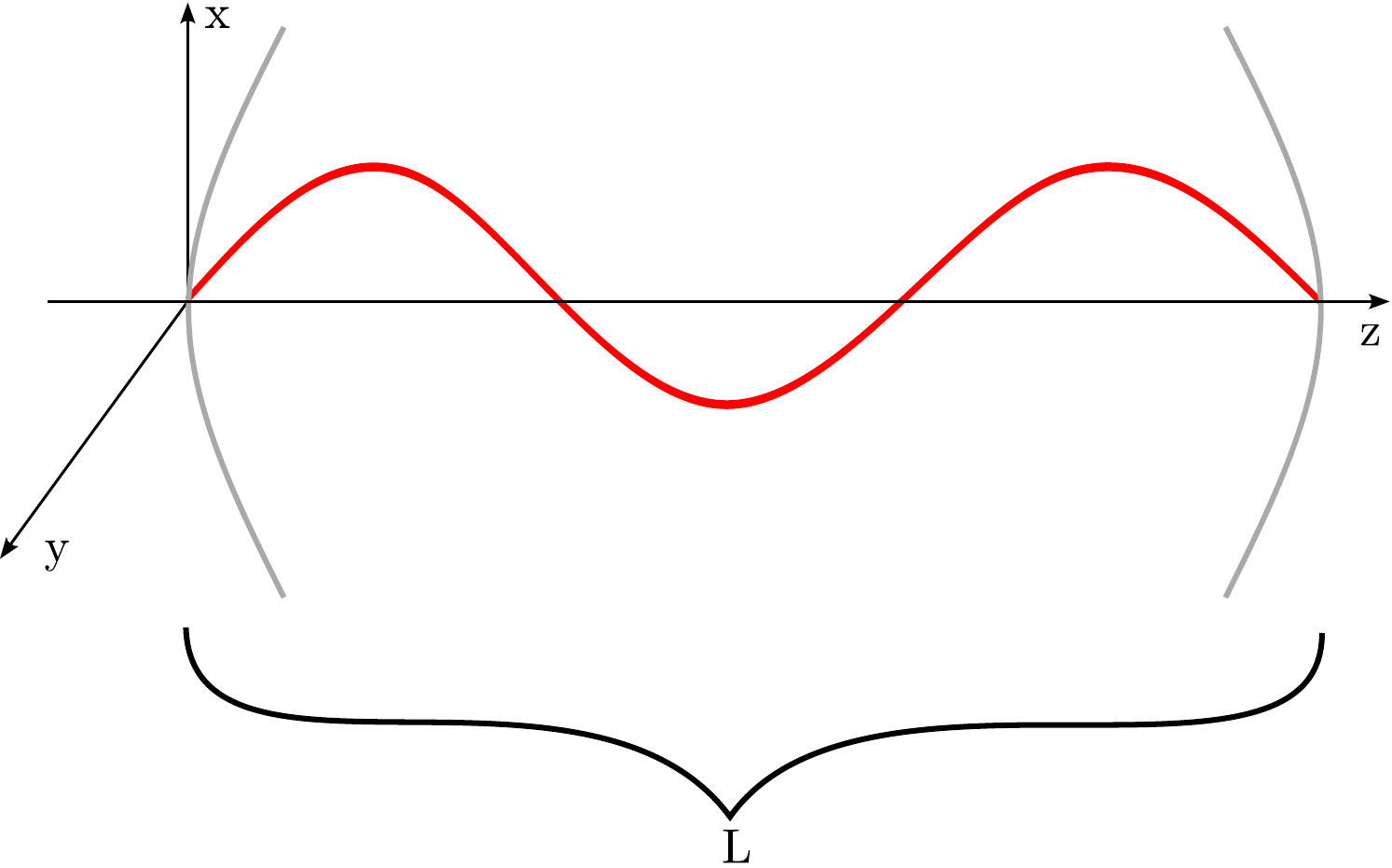}
  \caption{Standing wave inside a cavity with perfectly reflecting mirrors located at $z=0$ and $z=L$. The wave is polarised in the x-direction.}
  \label{fig:cavity}
\end{figure}

\noindent where $\rho(t)$ = $\frac{dq}{dt}$ is the canonical momentum and $\mu_{0}$ is the permeability of free space. In this context the Hamiltonian ($H$), i.e. the total energy, of this single mode field is given by

\begin{equation}\label{eqt:cla_H}
\begin{aligned}
    H &= \frac{1}{2}\int dV[\epsilon_{0}\mathbf{E}^{2}(\mathbf{r},t) + \frac{1}{\mu_{0}}\mathbf{B}^{2}(\mathbf{r},t)] \\
      &= \frac{1}{2}\int dV[\epsilon_{0}{E_{x}^{2}}(z,t) + \frac{1}{\mu_{0}} {B}_{y}^{2}(z,t)].
\end{aligned}
\end{equation}

With (\ref{eqt:e_field}) and  (\ref{eqt:cla_mag}), (\ref{eqt:cla_H}) can be written in terms of $q$ and $\rho$ as

\begin{equation}
    H = \frac{1}{2}(\rho^{2}+\omega^{2}q^{2}).
\end{equation}

This is simply a harmonic oscillator of unit mass. To quantise, the correspondence principle can be invoked whereby the canonical variables $q$ and $\rho$ take the form of their operator equivalents $\hat{q}$, $\hat{\rho}= i\frac{\partial}{\partial\hat{q}}$ and satisfy the commutation relation $[\hat{q},\hat{\rho}] = i\hat{I}$. In turn, the electric and magnetic fields become operators ($E_x(z,t) \rightarrow \hat{E}_{x}(z,t)$, $B_{y}(z,t) \rightarrow \hat{B}_y(z,t)$)  and we arrive at the quantum harmonic oscillator

\begin{equation}
    \hat{H} = \frac{1}{2}(\hat{\rho}^2+\omega^2\hat{q}^2).
\end{equation}

$\hat{q}$ and $\hat{\rho}$ are used to define what are termed the annihilation ($\hat{a}$) and creation ($\hat{a}^\dagger$) operators. Mathematically, these operators describe the creation and annihilation of a quanta of energy (for example, a photon) and although unphysical are mathematically useful.

\begin{equation}\label{eqt:anni}
\begin{aligned}
    &\hat{a} = \frac{1}{\sqrt{2\hbar\omega}}(\omega \hat{q}+i\hat{\rho}) \quad ;
    &\hat{a}^\dagger = \frac{1}{\sqrt{2\hbar\omega}}(\omega \hat{q}-i\hat{\rho}).    
\end{aligned}
\end{equation}

using $\hat{a}$ and $\hat{a}^\dagger$, the electric and magnetic field operators can be re-written as 

\begin{equation}
\begin{aligned}
    &\hat{E}_x(z,t) = \mathcal{E}_0(\hat{a}e^{-i\omega t}+\hat{a}^\dagger e^{i\omega t})\sin(kz) \\
    &\hat{B}_y(z,t) = -i\mathcal{B}_0(\hat{a}e^{-i\omega t}-\hat{a}^\dagger e^{i\omega t})\cos(kz)
\end{aligned}
\end{equation}

\noindent with $\mathcal{E}_0 = (\frac{\hbar \omega}{\epsilon_0V})^{\frac{1}{2}}$ and $\mathcal{B}_0 = (\frac{\mu_0}{k})(\frac{\epsilon_0\hbar\omega^3}{V})^{\frac{1}{2}}$. $e^{\pm i\omega t}$ represents the time dependency of the operator. We can now introduce the quadrature operators as,

\begin{equation}\label{eqt:quad_op}
\begin{aligned}
    &\hat{x} = \frac{1}{\sqrt{2}} (\hat{a} + \hat{a}^\dagger) \quad ;
    &\hat{p} = \frac{1}{i\sqrt{2}} (\hat{a} - \hat{a}^\dagger),
\end{aligned}
\end{equation}

\noindent where $\hat{a}(t=0) = \hat{a}$. $E_x(z,t)$ can then be recast as

\begin{equation}\label{eqt:quad}
    \hat{E}_x(t) = \sqrt{2}\epsilon_{0} \sin(kz)[\hat{x}\cos(\omega t)+\hat{p}\sin(\omega t)].
\end{equation}

From (\ref{eqt:quad}), it is clear that $\hat{x}$ and $\hat{p}$ are associated with field amplitudes oscillating $\frac{\pi}{2}$-radians out of phase. Further, they are essentially the position and momentum operators obtained from (\ref{eqt:anni}) but scaled to be dimensionless and can therefore be viewed in terms of a phase space. 

The annihilation and creation operators and thus the quadratures can be made multi-modal by assuming a mode number $k$ with $k=1...,n$. Therefore 

\begin{equation}
\begin{aligned}
    &\hat{x}_k = \frac{1}{\sqrt{2}} (\hat{a}_k + \hat{a}_k^\dagger) \quad ;
    &\hat{p}_k = \frac{1}{i\sqrt{2}} (\hat{a}_k - \hat{a}^\dagger_k)
\end{aligned}
\end{equation}

\noindent which obey the commutation relation $[\hat{x}_k, \hat{p}_k] = i\delta_{kl}$. A general, rotated quadrature can be defined as

\begin{equation}
\begin{aligned}
    \hat{\mathfrak{q}}_k(\theta) &= \frac{1}{\sqrt{2}}(\hat{a}_ke^{i\theta} + \hat{a}_k^\dagger e^{i\theta}) \\
    &= \cos(\theta)\hat{x}_k + \sin(\theta)\hat{p}_k
    \end{aligned}\label{eqt:gen_quad}
\end{equation}

\noindent with $[\hat{\mathfrak{q}}_k(\theta),\hat{\mathfrak{q}}_l(\alpha)] = i\delta_{kl}\sin(\theta-\alpha)$. The continuous nature of the quadrature operators can be shown by considering their eigenstates. The operators \( \hat{x} \) and \( \hat{p} \) satisfy the eigenvalue equations\footnote{Quantum mechanics uses the formalism of linear algebra. The notation $\ket{\cdot}$ is called a \textit{ket}, representing a column vector in a complex vector space. Its Hermitian conjugate, $\bra{\cdot}$, is called a \textit{bra}, and represents the corresponding row vector (the complex conjugate transpose of the ket). The inner product of a bra and a ket, written $\langle \cdot | \cdot \rangle$, yields a complex number. The outer product $\ket{\cdot}\bra{\cdot}$ forms an operator. An expectation value of an operator $\hat{B}$ in a state $\ket{a}$ is denoted $\bra{a} \hat{B} \ket{a}$.
}:

\begin{equation}\label{eqt:quad_eigen}
\hat{x} \, |x\rangle = x \, |x\rangle, \qquad \hat{p} \, |p\rangle = p \, |p\rangle
\end{equation}

\noindent where $x, p \in \mathbb{R}$. These eigenstates form a continuous basis over the real line:

\begin{equation}
\int_{-\infty}^{\infty} |x\rangle \langle x| \, dx = \hat{I}, \qquad \langle x | x' \rangle = \delta(x - x')
\end{equation}

\noindent and similarly for \( |p\rangle \). The real-valued spectra of \( \hat{x} \) and \( \hat{p} \) imply that measurement outcomes span a continuum, and therefore these observables are referred to as \emph{continuous variables}.
The usefulness of the quadratures lies in the fact that they correspond to measurable physical observables, and can be readily accessed through homodyne detection. Before introducing homodyne detection and why feedforwarding is ultimately required in \gls{cv} \gls{mbqip}, it is imperative we first introduce the Wigner function and Gaussian states.

\subsubsection{The Wigner Function and Gaussian States}\label{subsec:Wig_Gauss}

In quantum mechanics, physical systems such as the electromagnetic field can be described by a matrix known as the density matrix, denoted by 
$\hat{\varrho}$, which encodes all statistically accessible information about the quantum state. Depending on these statistical properties, the density matrix can take various forms. A useful way to visualize these different forms is through the Wigner function, a quasi-probability distribution that represents quantum states in phase space \cite{Wigner1932}. The Wigner function is termed a quasi-probability because, unlike classical probability distributions, it can take on negative values. The presence of such negativity is a hallmark of what is termed a non-classical state of light. It thus provides a bridge between the abstract density matrix and a more intuitive, phase-space picture of the quantum state.

\begin{equation}\label{eqt:wigner}
    W(\mathbf{x,p}) = \frac{1}{(2\pi)^{n}}\int_{\mathbb{R}^n}\bra{\mathbf{x}+\frac{1}{2}\mathbf{q}}\hat{\varrho}\ket{\mathbf{x}-\frac{1}{2}\mathbf{q}}e^{i\mathbf{p\cdot\mathbf{q}}}d\mathbf{q}.
\end{equation}

Equation (\ref{eqt:wigner}) shows the Wigner function is expressed in terms of the real-valued variables $\mathbf{x}$ and $\mathbf{p}$, which correspond to the eigenvalues of the quadrature operators $\hat{x}$ and $\hat{p}$ introduced in (\ref{eqt:quad_eigen}). The variable $\mathbf{q}$ is a relative displacement vector and represents a spatial shift of $\mathbf{x}$, the midpoint. The term $\bra{\mathbf{x}+\frac{1}{2}\mathbf{q}}\hat{\varrho}\ket{\mathbf{x}-\frac{1}{2}\mathbf{q}}$ encodes the off-diagonal elements corresponding to the quantum coherence between the two points $\mathbf{x}+\frac{1}{2}\mathbf{q}$ and $\mathbf{x}-\frac{1}{2}\mathbf{q}$. The Wigner function essentially Fourier-transforms these off-diagonal elements with respect to $\mathbf{q}$ mapping the quantum state's coherence into the momentum variable $\mathbf{p}$ and therefore phase-space. 

\begin{figure}[t]
  \centering
  \includegraphics[width=\linewidth]{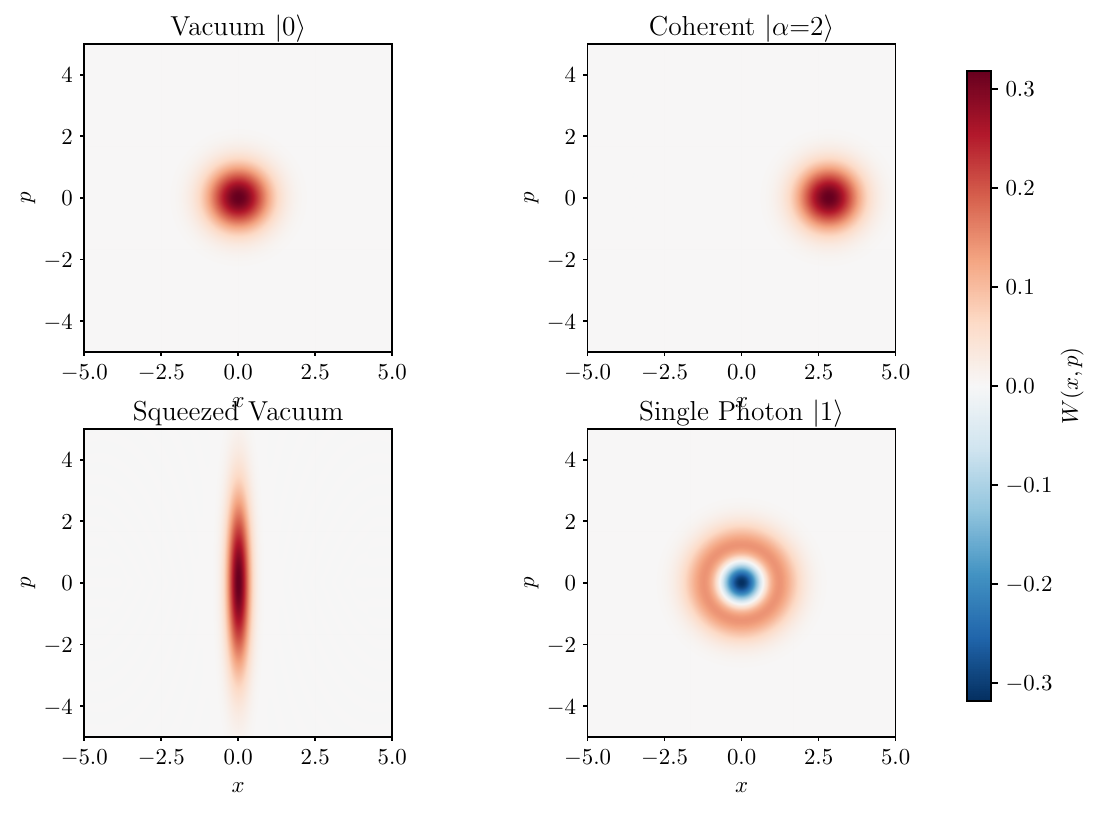}
  \caption{Wigner functions for vacuum, coherent ($\alpha=2$), squeezed vacuum, and single-photon states. The single-photon state is non-Gaussian and has a negative Wigner function.}
  \label{fig:wigner-panel}
\end{figure}

Gaussian states are those states for which the Wigner function is a (multivariate) Gaussian function\cite{bohr-brask}. They play a crucial role in \gls{mbqip} and are completely described by the expectation values (first moment) and covariance (second moment) of the quadrature operators introduced in (\ref{eqt:quad_op}). That is $\bar{\mathbf{r}}=\langle \hat{\mathbf{r}} \rangle$ and $
\boldsymbol{\sigma} = \frac{1}{2} \left\langle \{ \Delta \hat{\mathbf{r}}, (\Delta \hat{\mathbf{r}})^T \} \right\rangle = \frac{1}{2} \left\langle \hat{\mathbf{r}} \hat{\mathbf{r}}^T + (\hat{\mathbf{r}} \hat{\mathbf{r}}^T)^T \right\rangle - \bar{\mathbf{r}} \bar{\mathbf{r}}^T
$ where $\mathbf{\hat{r}}$ is a column vector of the quadratures, given, in what is termed $xxpp$-ordering, as:

\begin{equation}\label{eqt:disp_vector}
    \mathbf{\hat{r}} = \begin{bmatrix}
\hat{x}_1 &
\hat{x}_2 &
\cdots &
\hat{x}_n&
\hat{p}_1 &
\hat{p}_2 &
\cdots &
\hat{p}_n\\
\end{bmatrix}^T
.
\end{equation}

The symbol $\left\langle \cdot  \right\rangle$ is termed the expectation value and represents the average outcome of a measurement if that measurement were performed many times on identically prepared quantum systems. $\boldsymbol{\sigma}$ is termed the covariance matrix of a quantum state $\hat{\varrho}$ and has elements

\begin{equation}
\sigma_{ij} = \frac{1}{2} \left\langle \hat{r}_i \hat{r}_j + \hat{r}_j \hat{r}_i \right\rangle - \left\langle \hat{r}_i \right\rangle \left\langle \hat{r}_j \right\rangle .
\end{equation}

The most common Gaussian states in optics are vacuum states (representing the absence of photons), coherent states (laser light) and  squeezed states (were one quadrature values' variance can be less than that of the vacuum). Their Wigner functions are depicted in Fig. \ref{fig:wigner-panel} alongside the non-Gaussian single photon state. It is clear at this point the importance of the quadratures in \gls{cv} \gls{mbqip} which of course require some way of being measured. This is achieved via homodyne detection.

\subsubsection{Gaussian Operations, Homodyne Detection, and Displacement}

\begin{figure}[!ht]
\centering
\includegraphics[width=\linewidth]{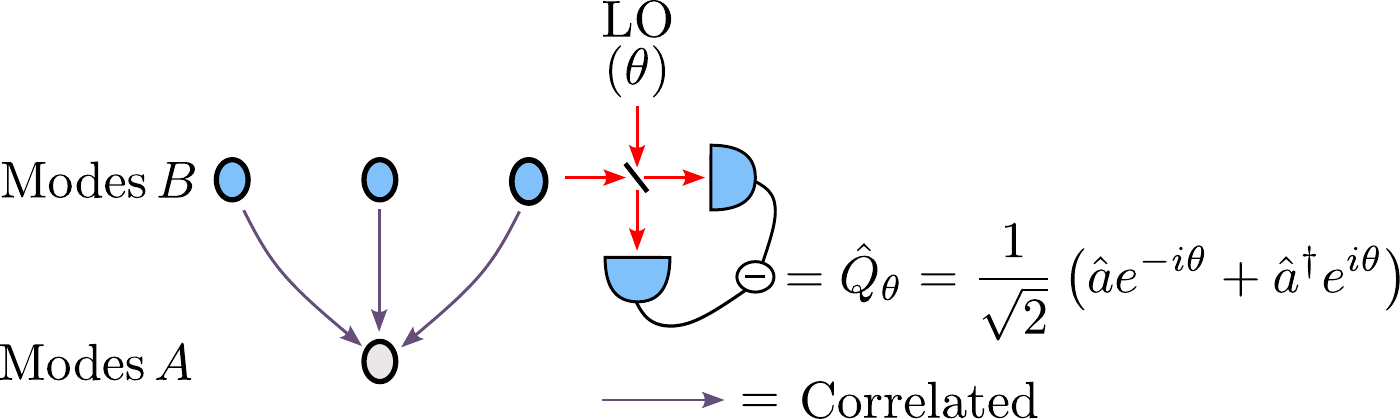}
\caption{Concept of a joint system in the Gaussian formalism. Modes B are correlated with Mode A. Measuring Modes B via homodyne detection leads to a displacement in Mode A. This can be viewed as an error that must be corrected for. This happens via feedforwarding.}
\label{fig:joint_sys}
\end{figure}

In quantum mechanics, an operation refers to any transformation of a quantum state. For Gaussian states, fully described by their first moments $\bar{\mathbf{r}}$ and covariance matrix $\boldsymbol{\sigma}$, Gaussian operations are those that preserve the Gaussian character of the state. Unitary Gaussian operations (e.g., phase rotations, \gls{bs}, and squeezing) act as follows:

\begin{equation}
\bar{\mathbf{r}} \rightarrow \mathbf{S}\bar{\mathbf{r}} + \mathbf{d}, \quad \boldsymbol{\sigma} \rightarrow \mathbf{S}\boldsymbol{\sigma}\mathbf{S}^\dagger,
\end{equation}

\noindent where $\mathbf{S}$ is a symplectic matrix satisfying $\mathbf{S}\Omega\mathbf{S}^\dagger = \Omega$, with

\begin{equation}
\Omega = \bigoplus_{k=1}^n
\begin{pmatrix}
0 & 1 \\
-1 & 0
\end{pmatrix}.
\end{equation}

\begin{figure}[t!]
\centering
\includegraphics[width=\linewidth]{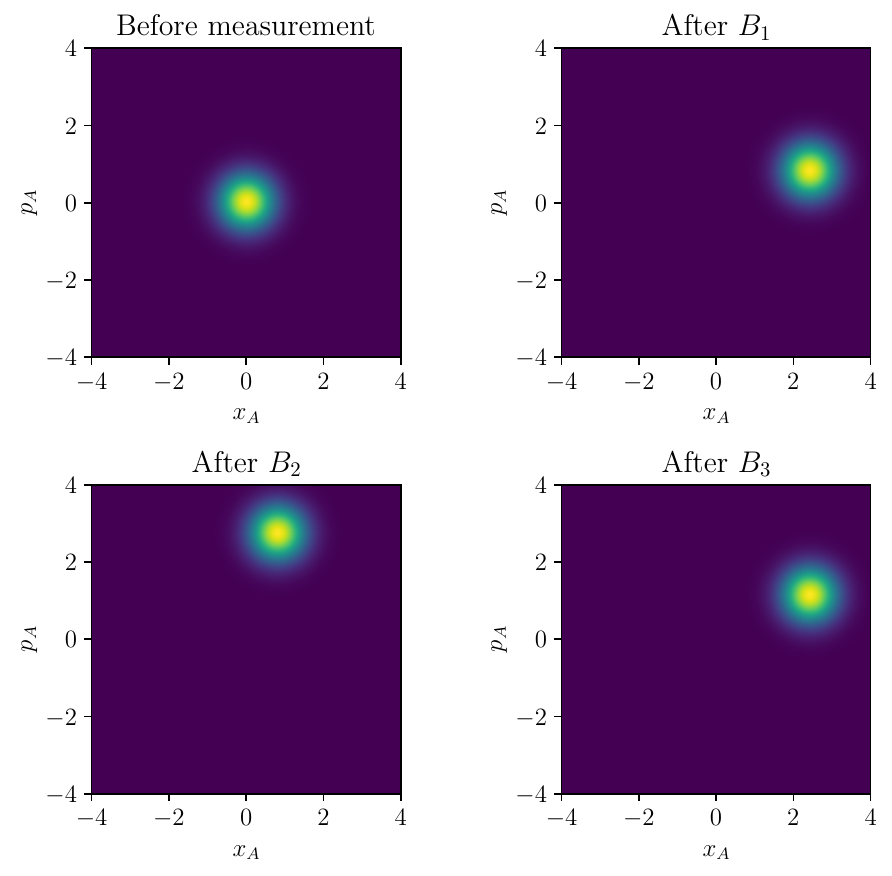}
\caption{Wigner functions showing subsystem $A$ before and after $\hat{x}$-measurement on subsystem $B$. Homodyne detection displaces $A$ in phase space, requiring real-time correction.}
\label{fig:aff_HD}
\end{figure}

Non-unitary Gaussian operations, such as homodyne detection, are irreversible and not described by symplectic matrices alone. Homodyne detection projects a mode onto a quadrature eigenstate, collapsing the global state and conditioning the remaining subsystem. Consider a $n$-mode Gaussian state where the last mode undergoes a homodyne measurement (Fig \ref{fig:aff_HD}). We call this mode $B$ and all non-measured modes $A$. We therefore have a joint system with a covariance matrix and mean vector:

\begin{align}
\boldsymbol{\sigma} &= \begin{bmatrix}
\boldsymbol{A} & \boldsymbol{C} \\
\boldsymbol{C}^T & \boldsymbol{B}
\end{bmatrix}, &
\bar{\mathbf{r}} &= \begin{bmatrix}
\boldsymbol{a} \\
\boldsymbol{b}
\end{bmatrix}.\label{eqt:dis_cov_matr}
\end{align}

Here $\boldsymbol{A}\,(\boldsymbol{B}) =\boldsymbol{\sigma}_A\,(\boldsymbol{\sigma}_B)$, $\boldsymbol{C}$ encodes correlations between $\boldsymbol{A}$ and $\boldsymbol{B}$ and  $\boldsymbol{a}\,(\boldsymbol{b}) = \bar{\mathbf{r}}_A \, (\bar{\mathbf{r}}_B)$. A homodyne measurement of $\hat{x}$ on mode $B$ leads to a conditional update of subsystem $A$. The post-measurement covariance matrix becomes\cite{bohr-brask}

\begin{equation}
\boldsymbol{\sigma}_A = \boldsymbol{a} - \frac{1}{B{11}} \boldsymbol{C}\boldsymbol{\Pi} \boldsymbol{C}^T,
\end{equation}

\noindent and the conditional mean is shifted by

\begin{equation}\label{eqt:dis_new}
\bar{\mathbf{r}}_A = \boldsymbol{a} - \frac{1}{B{11}} \boldsymbol{C}\boldsymbol{\Pi}(\boldsymbol{b} - \boldsymbol{s}),
\end{equation}

\noindent where $\boldsymbol{\Pi} = \begin{bmatrix} 1 & 0 \\ 0 & 0 \end{bmatrix}$ and $\boldsymbol{s} = \begin{bmatrix} s \\ 0 \end{bmatrix}$ contains the measurement result. The displacement term in Eq.~\eqref{eqt:dis_new} is deterministic, conditioned on $s$, and must be corrected via feedforward to restore the unmeasured system to its intended logical state. This displacement reflects the fact that, due to correlations between subsystems ($\boldsymbol{C}\neq0$), the measurement on mode $B$ reveals partial information about mode $A$, updating our best estimate of its state and resulting in a conditional shift of its mean vector. Figure \ref{fig:aff_HD} illustrates this conditional displacement effect in phase space.

The measurement outcome $s$ follows a Gaussian probability distribution centred at $b_1$ with variance $B_{11}$:

\begin{equation}\label{eqt:wig_marg_new}
P_{\hat{x}}(s) = \frac{1}{\sqrt{2\pi B_{11}}} \exp\left[-\frac{(s - b_1)^2}{2B_{11}}\right].
\end{equation}

In summary, homodyne detection acts as a Gaussian measurement channel that collapses and displaces the state, necessitating a feedforward correction proportional to the measurement result. Implementing this correction in real-time is the core motivation behind the \gls{ff}-system in CV MBQIP.

\subsubsection{Measurement-Based Quantum Information Processing}

\begin{figure}[!ht]
    \centering    
    \includegraphics[width=0.7\linewidth]{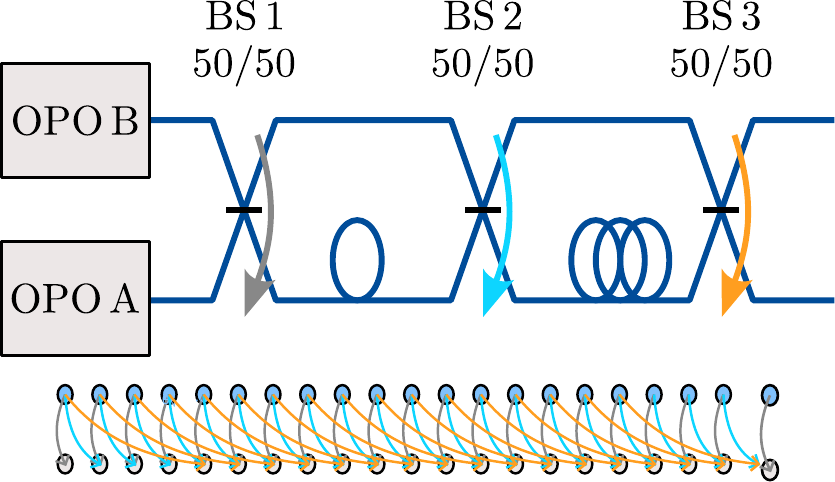}
    \caption{Overview of how a cluster state is generated, here using time multiplexing via delay lines, and multiple beamsplitters. This has the effect of implementing control-Z gates and entangling multiple modes.}
    \label{fig:MB-gbs}
\end{figure}

In contrast to the circuit model of quantum computing, where logic gates are applied sequentially to quantum states via optical components, \gls{mbqip} shifts the burden of computation onto a pre-generated highly entangled resource known as a cluster state. For continuous-variable (CV) systems, this cluster state is composed of many optical modes, each prepared in a squeezed vacuum state, and entangled using the CV controlled-Z ($\hat{C}_Z$) gate via a \textit{fixed} number of linear optical components (Fig \ref{fig:MB-gbs}). The CV controlled-Z gate entangles\footnote{Entanglement is essentially a fancy word for 'very highly correlated'. When entangled, modes are more correlated then is possible 'classically'.} modes by shifting each mode's momentum quadrature by the position quadrature of the other,

\begin{equation}
\hat{C}_Z :
\begin{cases}
\hat{q}_1 \rightarrow \hat{q}_1 \\
\hat{p}_1 \rightarrow \hat{p}_1 + \hat{q}_2 \\
\hat{q}_2 \rightarrow \hat{q}_2 \\
\hat{p}_2 \rightarrow \hat{p}_2 + \hat{q}_1,
\end{cases}
\end{equation}

\noindent and is termed two-mode entanglement. Each mode in the cluster state represents a computational "wire". Information flows through the cluster not by propagating qubits, but by performing measurements on each mode in sequence. The choice of measurement basis, i.e., the choice of $\theta$ in (\ref{eqt:gen_quad}), determines which quantum gate (single-mode unitary), is effectively applied to the logical state. The entanglement allows this flow of information, termed 'teleportation'. Numerous measurements on particular modes are typically necessary to carry out one gate\cite{rauss_cnot_gate}. This makes \gls{mbqip} measurement-driven and classically adaptive.

Mathematically, the measurement collapses the entangled subsystem into a conditional Gaussian state, similar to Eq. (\ref{eqt:dis_new}). The resulting displacement in phase space thus depends linearly on the measured value $s$, and must be undone by applying a counter-displacement to a downstream mode using an optical modulator. Secondly, unlike in qubit based MBQC, CV MBQC measures real numbers instead of binary 0's and 1's, therefore requiring updates of the form:

\begin{equation}\label{eqt:inner_product_calc}
    x_{out} = \boldsymbol{A}\cdot \boldsymbol{s}
\end{equation}

\noindent where $\boldsymbol{A}$ is a matrix that can be known in advance, $x_{out}$ is a scalar correction sent to a modulator and $\boldsymbol{s}$ is a vector of multiple measurements.

This structure-entangled resource, adaptive measurement, and feedforward correction-makes CV MBQIP highly scalable. It also highlights why real-time processing is essential: if the displacement corrections are delayed, the quantum state evolves incorrectly, degrading the computation. This in turn imposes strict latency conditions on the classical electronics.

\section{FPGA Based Fast Feedforward System}

Now we understand the motivation behind having a fast feedforward system we can introduce it. The most relevant
equation as far as the FF-system is concerned is Eq. \eqref{eqt:inner_product_calc}.  Alas, the \gls{fpga} is designed to  carry out an inner-product calculation between the rows of pre-programmed matrices (termed $A$-vectors) and a vector of measurement results (termed $m$-vectors); and for this output to be translated to polar co-ordinates to allow for movement around phase space. Further it contains multiple entities that are required for experimental control which will be discussed below. It operates with a 250 Mhz clock. The system is built around the AMD ZCU102 \gls{fpga} development board and includes a 1 GHz 12-bit ADC (TI ADS5400), a 250 MHz 12-bit DAC (TI DAC5662), and a custom-built \gls{fdb}. All entities are implemented in VHDL, with freeRTOS as the operating system. 

\subsection{System Overview and Data Flow}

\begin{figure*}[t!]
\centering
\includegraphics[width=0.8\linewidth]{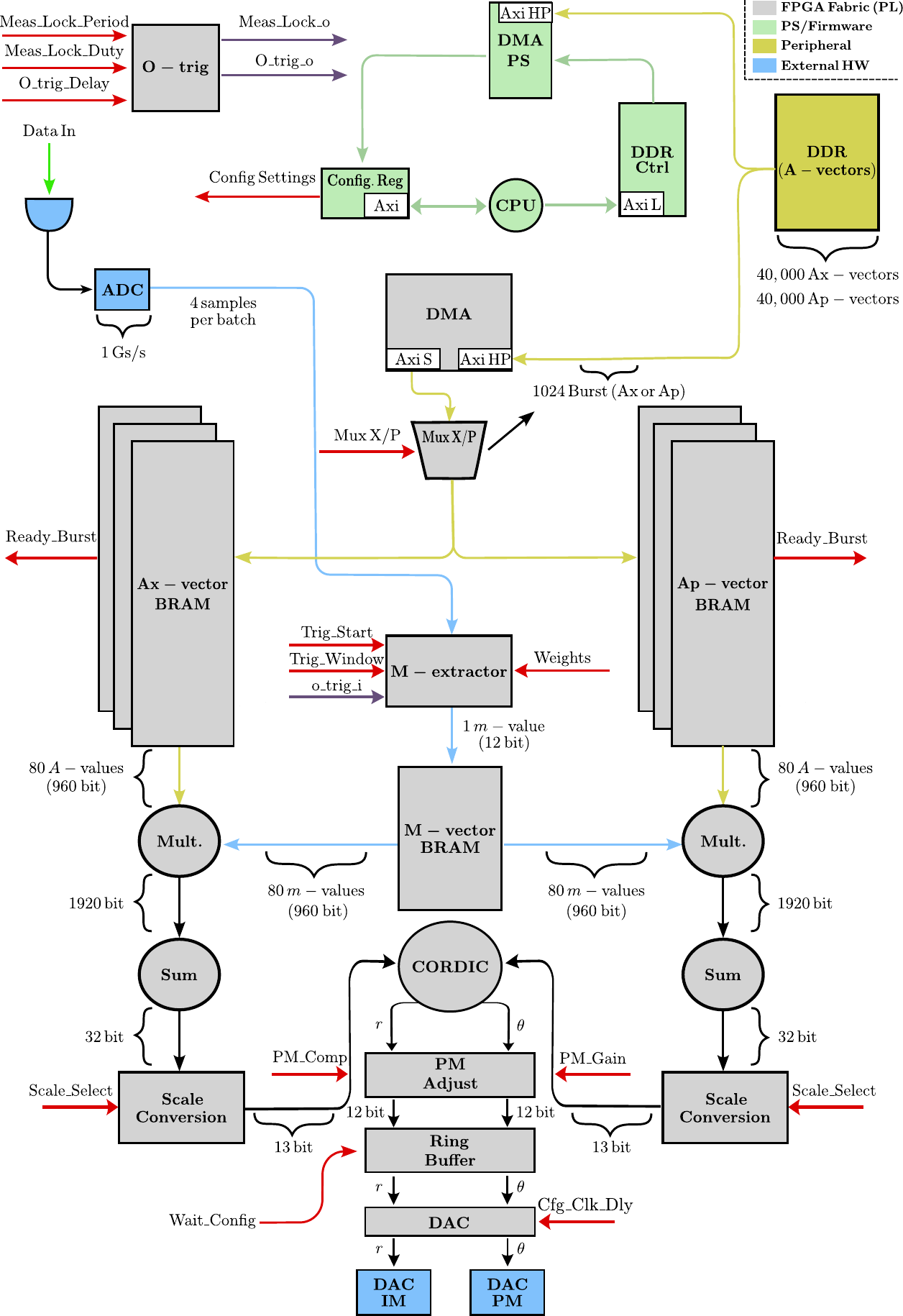}
\caption{Feedforward System architecture and data flow. $A$-vectors are stored in DDR (top right), and sent to the Ax/Ap-vector BRAM when required. An optical pulse is measured by a high quantum efficiency detector (top left). This is sampled by the ADC and 1 $m$-value is created via the M-extractor. A summer and multiplier carry out the inner-product calculation between the appropriate $A$-vector and the $m$-vector. A user configurable entity, Scale Conversion, is used to select which 13 bits of the resultant 32-bit value is to be sent to the CORDIC, which carries out the rectangular to polar coordinate translation. Another user configurable entity, Ring Buffer, can be used to delay the output if required. The 12-bit result of the calculation is sent to two DACs, one which drives an \gls{im} and one which drives a \gls{pm}.}
\label{fig:ff_arch}
\end{figure*}

\begin{figure}[t]
    \centering

    \subfloat[]{
        \includegraphics[width=0.9\linewidth]
        {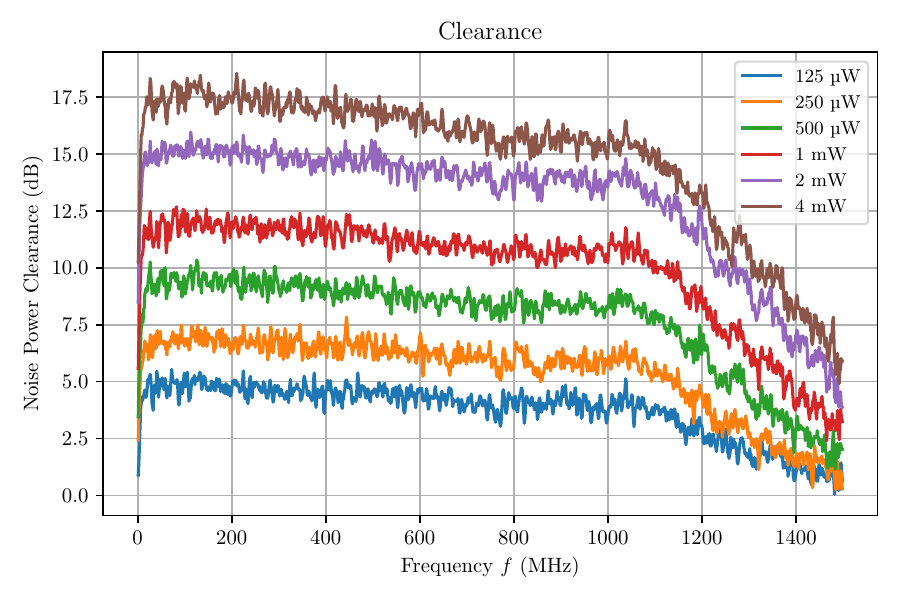}
        \label{fig:sub1}
    }

    \subfloat[]{
        \includegraphics[width=0.9\linewidth]
        {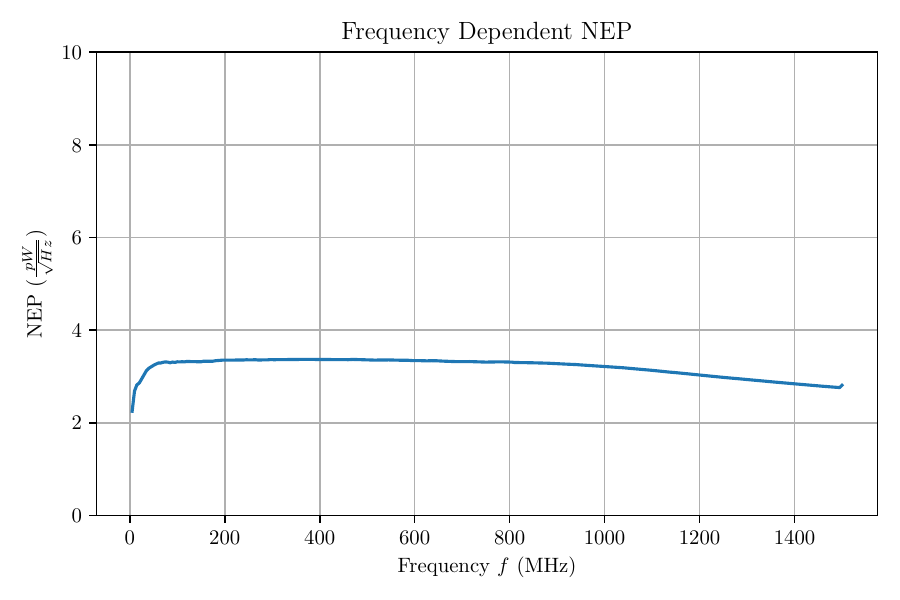}
        \label{fig:sub2}
    }

    \caption{
    Clearance and noise equivalent power (NEP) of the homodyne detector.
    The clearance shows less than 3 dB variation across the signal band,
    with a clearance of $>16$ dB at 400 MHz and 15 dB at 1 GHz for
    4 mW \gls{lo}. The NEP is approximately
    $3\,\mathrm{pW}/\sqrt{\mathrm{Hz}}$ across the signal band.
    }
    \label{fig:HD_combined}
\end{figure}

Figure \ref{fig:ff_arch} shows the data flow of the system and for our purpose will act as a reference block diagram. Optical signals are measured via a low noise (NEP $\approx$ 3 $\frac{pW}{\sqrt{Hz}}$), fully fibre-based \gls{hqe} (>95\%) \gls{hd}, with a clearance of 15 dB at 1 GHz with 4 mW \gls{lo}\footnote{2 mW on each diode.} (Fig. \ref{fig:HD_combined}). The system is designed to operate with a 10 Mhz pulsed laser, thus a new pulse arrives at the \gls{hd} every 100 ns. The output is sent to an ADC. This samples the data at 1 GS/s, and produces 1 $m$-vector value ($m$-value) via the M-extractor, a mode function with user defined weights. The $m$-value is stored in a BRAM termed the M-vector BRAM. It is suggested from theory that when the optical input is a cluster state, one measured mode displaces 80 non-measured modes, therefore the M-vector BRAM is designed as a shift register, capable of holding 80 12-bit values. Further, the calculation engine only starts once the $m$-vector contains 80-values. The calculation requires approximately 160,000 $A$-vectors (80,000 $A_x$-vectors, 80,000 $A_p$-vectors). Each $A$-vector also has 80 values of 12 bits each. Due to the large number, these are stored off the FPGA in DDR4 memory (4 GB) and read into the Ax/Ap-vector BRAM via a \gls{dma} module. They are written to DDR via Ethernet. 

Making use of Fig. \ref{fig:ff_arch} the calculation engine proceeds as follows, were we assume all three BRAM's are full and we focus on the $\hat{x}$-quadrature side of the calculation:

\begin{enumerate}
  \item The first 960-bit vectors from the Ax-BRAM and M-vector BRAM are each parsed into 80 signed integers of 12-bits inside the Multiplier entity. These vectors are multiplied element-wise producing an output vector consisting of 80 signed 24-bit products. This result is serialized into a 1920-bit bus, which is forwarded to the summation entity.
  \item The summer performs a pipelined summation of the 80 individual 24-bit products to produce a single 32-bit value representing the inner product scalar.
  \item The inner product result is passed to the Scale Conversion entity whose output is 13-bit. Based on a user configurable input, \textit{Scale Select}, the input is right-shifted. The entity sign-extends to preserve signedness and rounds to maintain numerical accuracy. The result is subsequently saturated to ensure it remains within the valid 13-bit signed range and is output to the CORDIC.
  \item The CORDIC turns this rectangular input into a polar output, specifically a magnitude ($r$). 
  \item The PM-Adjust entity compensates for any unwanted phase changes induced by the \gls{im} via \textit{PM-Comp}. Further it resizes the 13-bit input to 12 bits. It is this value that ultimately drives the \gls{im}. It is sent to the Ring Buffer.
  \item The Ring Buffer outputs its data after a user configurable number of clock cycles. 
  \item The DAC turns the digital signal to analog and outputs to the \gls{im} every 10 MHz.
\end{enumerate}

The above calculation is simultaneously duplicated for the $\hat{p}$-quadrature side. However, instead of a magnitude, the CORDIC outputs a \textit{signed} 13-bit phase ($\theta)$ that ranges from $-\pi$ to $\pi$. Further, inside the PM-Adjust entity, a user defined gain, \textit{PM-Gain}, can be set which is used to calibrate the \gls{pm} response. Before output, the data is offset by 1024, as the DAC cannot output negative values. 

\begin{figure}
\centering
\includegraphics[width=0.9\linewidth]{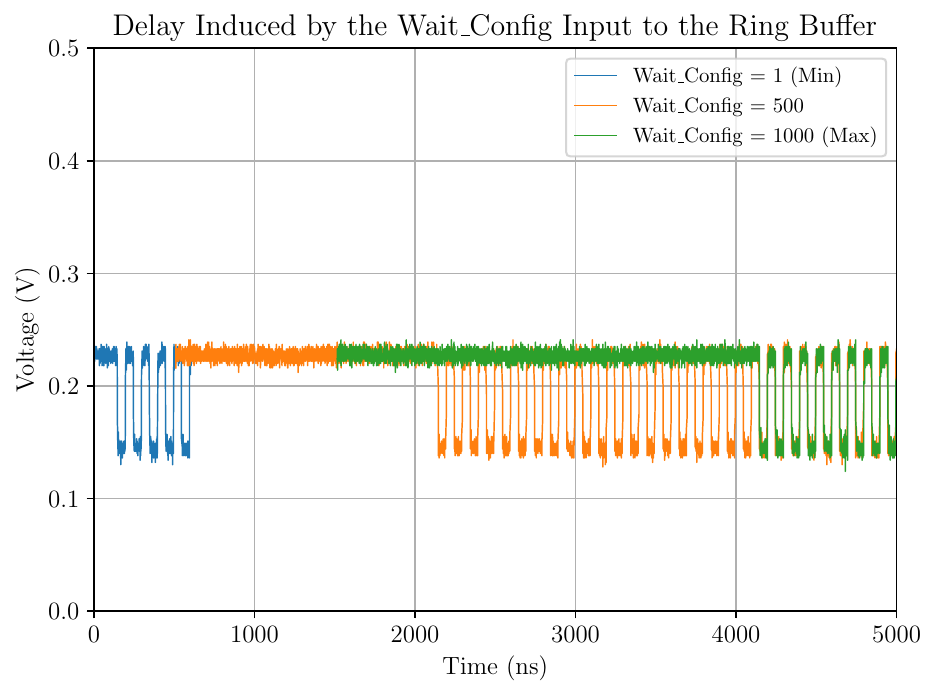}
\caption{The Ring Buffer can delay output from the DAC via a user configurable input \textit{Wait\_Config}, in steps of 4 ns. This is achieved by delaying the pulsing of the 10 MHz read enable signal for the specified amount of time. The max delay is approximately 4 µs.}
\label{fig:ring_buf}
\end{figure}

The CORDIC entity used is an AMD IP block\cite{amd_cordic}. The ring buffer is a dual-port ring buffer which can store up to 64 entries. Each entry contains a 24-bit word composed of concatenated $\hat{x}$ and $\hat{p}$ values. The buffer includes logic to prevent overflows and track near-full conditions. A state machine manages readout timing by waiting a user configurable number of clock cycles (via \textit{Wait Config}) after a write event before enabling output (Fig. \ref{fig:ring_buf}). This wait time allows synchronisation with pulses in the delay line if required, and can be set to 0 if no delay is necessary. The max delay is 4 µs. Once triggered, data is read sequentially from the buffer and made available to the DAC. A mathematical description of the calculation engine is given in appendix \ref{sec:calck_math}.

\subsection{Clock Planning}

\begin{figure}
\centering
\includegraphics[width=0.9\linewidth]{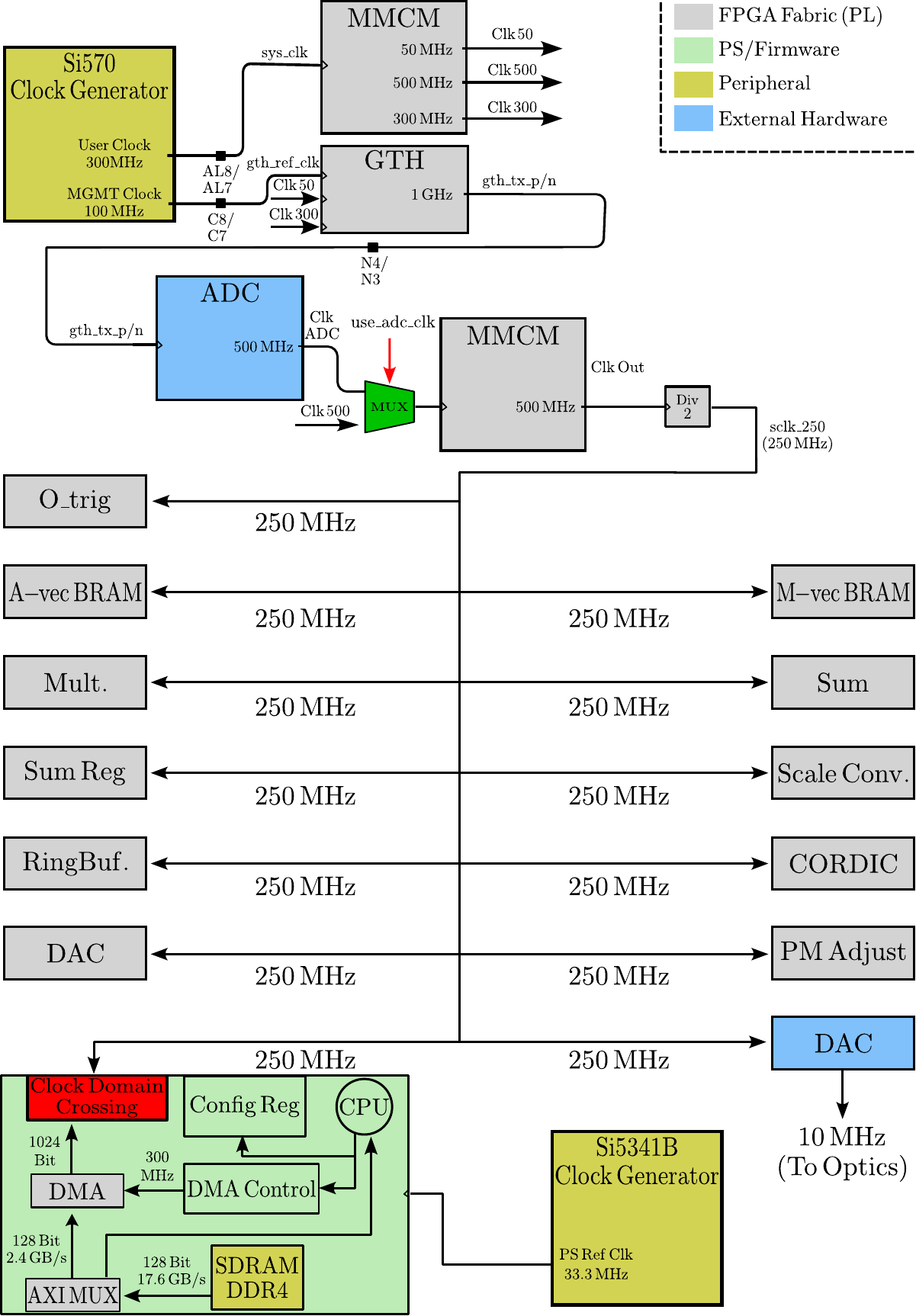}
\caption{Clock Plan. The onboard clock generator Si570 sends a reference clock to the \gls{gth} peripheral on the ZCU102 board. This is used to create the 1 GHz reference clock for the ADC. The ADC outputs a 500 MHz clock (Clk ADC) which is ultimately divided by 2 to create the 250 MHz fabric clock (sclk\_250). The top \gls{mmcm} is used to create necessary clocks for the GTH (Clk 50 and Clk 300, input not shown). Clk 500 is only used when the ADC clock (Clk ADC) is unavailable.}
\label{fig:clock_plan}
\end{figure}

As the TI ADS5400 ADC samples at 1 GS/s it requires a clock of 1 GHz. This can not be provided directly from a \acrlong{mmcm} (MMCM) module. It is seen from Fig. \ref{fig:clock_plan} that it is derived via the \acrlong{gth} block onboard the ZCU102.  The onboard oscillator Si570 generates a 300 Mhz reference which is divided by a \gls{mmcm} into 3 clock outputs, Clk 50, Clk300 and Clk 500. CLK 500 is only used when the other 500 MHz clock (ADC Clock) is unavailable. It is then used to clock a SERDES block and to produce the 250 MHz fabric clock (sclk\_250) via the second \gls{mmcm} instead. The \gls{gth} requires a stable free running clock for correct operation, which must be running before the \gls{gth} is configured\cite{amd_pg182_gth_port-des}. This is provided by Clk 50. Clk 300 is used for reset sequencing when the \gls{gth} module is reset.

The Si570 outputs a second reference clock (MGMT Clock 100 MHz) sent directly to the \gls{gth}. This is used by the \gls{gth}'s quad \gls{pll} to produce a 1 GHz clock output which is sent to the ADC. The ADC in turn outputs a 500 MHz clock (Clk ADC) which is subdivded in the second MMCM to give the 250 MHz clock used in the \gls{fpga} fabric (sclk\_250). The PS-side operates on a 300 MHz clock created via a PLL dedicated to the PS.

\subsection{Synchronisation and Experimental Control}

We now introduce the experimental control signals, along with the user-configurable inputs that govern system behavior. Two such inputs, \textit{Scale Select} and \textit{Wait Config}, were discussed earlier and are therefore not repeated here. All user-configurable parameters are stored in a configuration register (Config. Reg), which interfaces with the system’s control logic to dynamically adjust operating conditions. Register read and write operations are handled via a Python script that communicates over a Telnet connection, using the \gls{lwip} TCP/IP stack. The control signals are routed through a custom-built \gls{fdb} board, essentially a mezzanine card (FMC) equipped with level shifters, which interfaces with external experimental hardware. Refer to Fig. \ref{fig:ff_arch} for an overview of the signal flow and control architecture.

\subsubsection{Synchronisation Signals - O\_trig\_o and O\_trig\_i}

\begin{figure}
\centering
\includegraphics[width=0.8\linewidth]{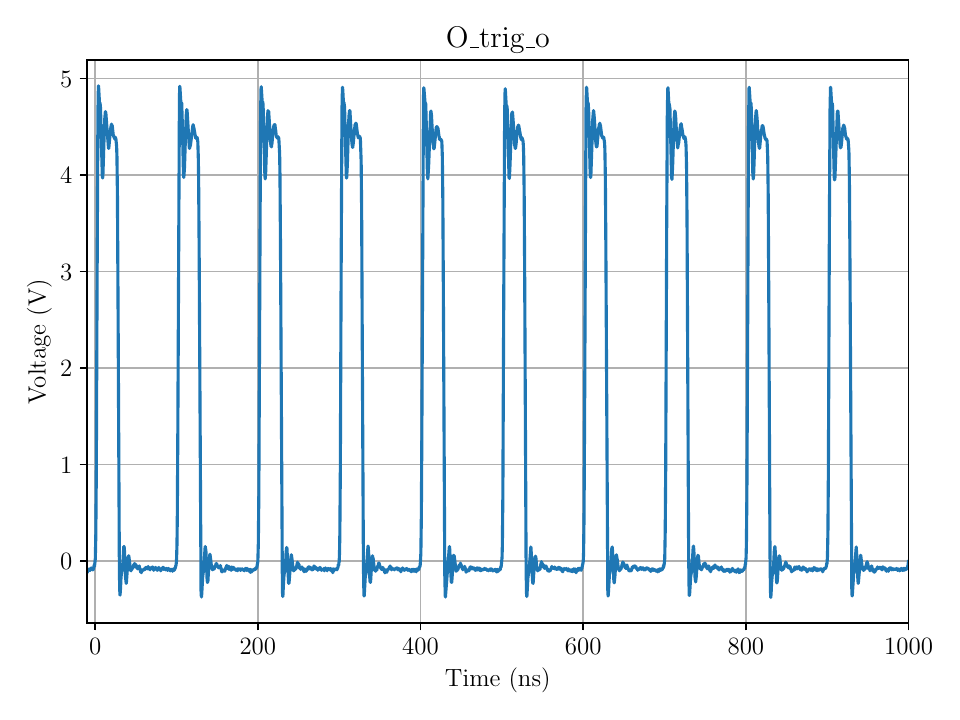}
\caption{O\_trig\_o is a 10 MHz signal used to synchronise the electronics and optics. Its output can be hastened with respect to O\_trig\_i}
\label{fig:o_trig}
\end{figure}

The \gls{ff}-system is designed to operate with a 10 MHz laser. In order to synchronise the optics with the electronics, it outputs a 10 MHz signal termed 'O\_trig\_o', i.e. optical trigger out, derived from the 250 MHz fabric clock via a counter. Its primary function is to drive a pulse generator used to create the optical pulses. As such, the feedforward system acts as the master clock. The M-extractor entity is directly synchronised to O\_trig\_o, via an internal signal O\_trig\_i, and and all other entities within the \gls{fpga} fabric (with the exception of the \gls{dma}) are synchronised indirectly through a series of Enable ports. These ports gate the flow of data into and out of each entity. O\_trig\_o can be hastened with respect to O\_trig\_i via a programmable offset parameter in the Otrig entity (\textit{O\_trig\_delay}) to compensate for fixed latencies in the optical path, such as those introduced by the pulse generator, cabling, and the ADC sampling process. This ensures that the arrival of optical pulses at the ADC is precisely aligned with the internal timing of the M-extractor, which begins processing data on the rising edge of O\_trig\_i. By fine-tuning this hastening, the system can guarantee that valid and correctly-timed data is available for downstream processing, thereby maintaining temporal coherence between the optical and electronic subsystems. O\_trig\_o can be set to output between 4 and 96 ns before O\_trig\_i in fabric clock-cycle steps (4 ns). The system outputs 2 O\_trig\_o signals (3.3 V and 5 V).

\begin{figure}
\centering
\includegraphics[width=0.9\linewidth]{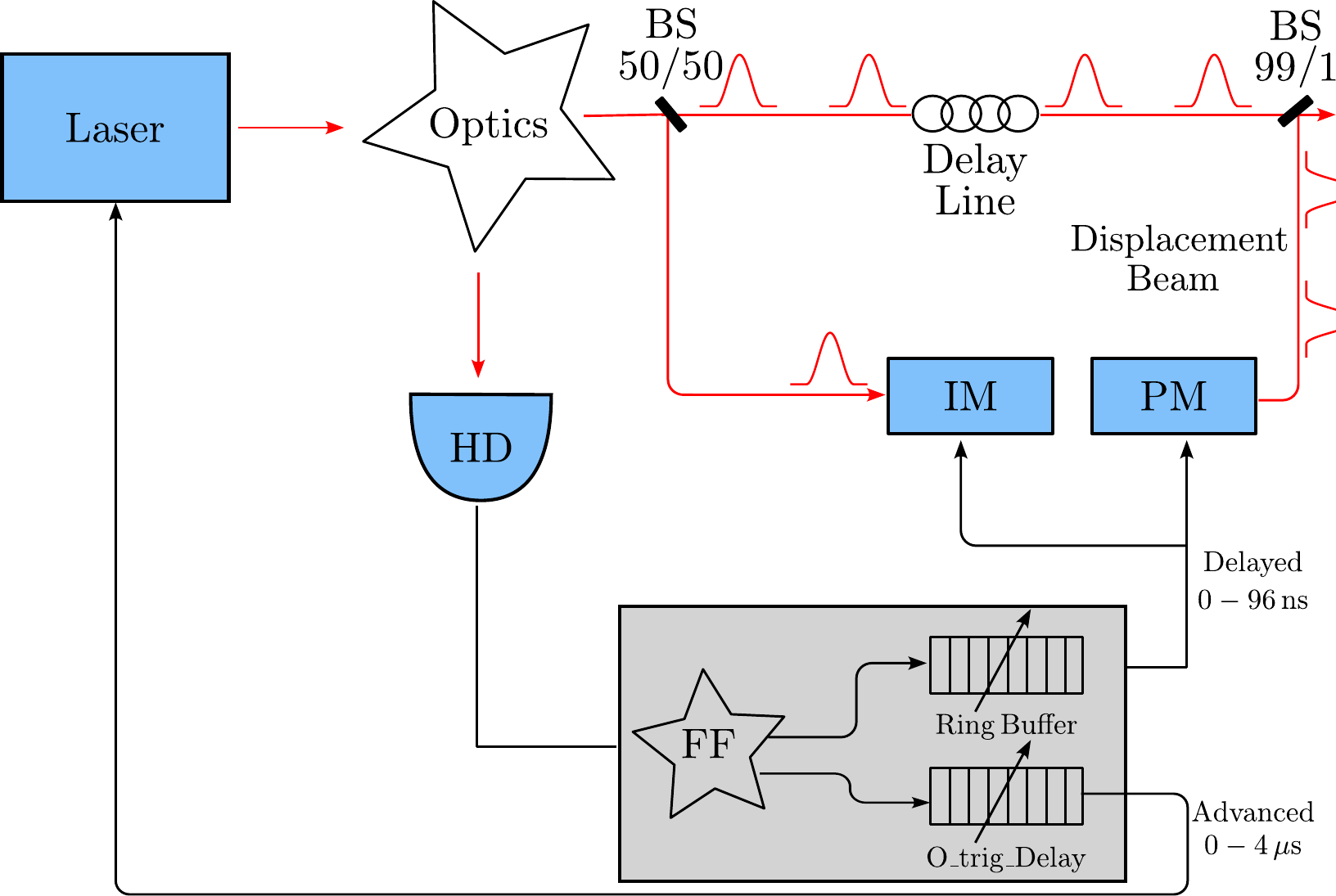}
\caption{The Ring Buffer delays the output data driving the IM and PM by 0–96 µs, and the O\_trig\_o signal controlling laser pulses by 0–96 ns. Both maintain temporal coherence between optics and electronics if required. The intensity and phase modulated displacement beam interferes with the delayed signal on a 99:1 beam splitter, implementing phase-space displacements.}

\label{fig:delay_params}
\end{figure}

\subsubsection{Measurement Signals - Meas\_Lock\_o}

\begin{figure}
\centering
\includegraphics[width=0.9\linewidth]{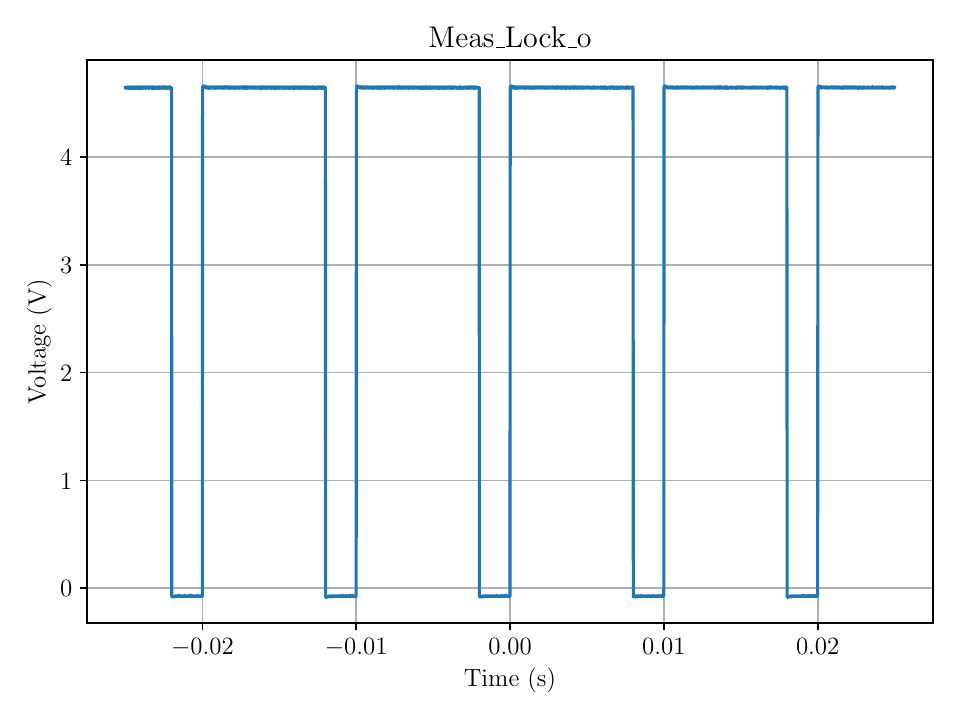}
\caption{Lock\_meas\_o is a configurable signal used to control the sample-hold scheme. It's default frequency is 100 Hz.}
\label{fig:lock_meas}
\end{figure}

In photonic experiments, numerous phase and cavity locks are typically required to ensure stable interference conditions and maintain coherence between optical paths. As an example, the delay line in the proposed experiments requires phase locking. An oft used approach is termed the sample-hold scheme\cite{casper_thesis}, where feedback loops are used for stabilisation of the optical system. This is typically achieved using the Pound–Drever–Hall technique\cite{pound_dh_technique}. In this approach, the optical system undergoes a 'sample' period during which the feedback loops and beams are activated to stabilize the system, followed by a 'hold' period during which they are turned off and measurement data is collected. The \gls{ff}-system is used to control this sample-hold scheme via the signal 'Meas\_Lock\_o'. The default frequency is 100 Hz and it is derived from O\_trig\_i. The user can configure the length of the sample period as required via \textit{Meas\_Lock\_Duty} i.e. by changing the duty cycle. Further the frequency  may be adjusted via \textit{Meas\_Lock\_Period} and has a range of 100 ns to approx. 6 s. 10 ms is the default and a typical value. There are 3 Meas\_Lock\_o outputs (two 5 V and 3.3 V).

\subsubsection{Data Deserialisation, the M-extractor and Weights}

In the \gls{ff}-system, high-speed digitized data from the ADC is first deserialised and then processed to extract relevant measurement values via the M-extractor\footnote{Technically in the design this is all achieved via one large entity simply called ADC.vhd i.e., the M-extractor is not a stand-alone entity as depicted in Fig \ref{fig:ff_arch}.}. The deserialised data is made available on a 48-bit wide bus clocked at 250 MHz. The data originates from the 12-channel, 1 GS/s ADC, where each channel transmits serial data at 1 Gbps. These are deserialised in a 1-to-4 scheme via a SERDES block: each of the 12 high-speed serial lines are converted into a 4-bit wide parallel word. As a result, the system receives a total of 48 bits (12 × 4) every 4 ns, synchronized to the 250 MHz system clock. 

Clock management is handled by a \gls{mmcm}. This \gls{mmcm} generates the required 250 MHz system clock either from CLK 500 or from ADC Clock as depicted in Fig. \ref{fig:clock_plan}. Further, ADC Clock is used in the SERDES block (or CLK500 when the ADC is unavailable). The phase of this clock is finely tuned to ensure correct timing alignment with the incoming ADC data.

The M-extractor is a mode filter, which acts as an electronic version of a \gls{tmf}, a mathematical construct that defines how quantum information is distributed over time in continuous-variable systems\cite{brecht_temporalmodes}. In \gls{cv}-optics, \gls{tmf} are typically implemented in post-processing and take the form

\begin{equation}\label{eqt:tmfunc}
    Var = \int w(t)X_i(t)dt
\end{equation}

\noindent whereby $var$ stands for variance and $w(t)$ are time-dependent weights. Essentially, mode functions are used to extract the relevant quadrature information from the data, by, for example, ignoring noise such as vacuum noise. Typically, such a process is undertaken in post-processing. The \gls{ff}-system allows for this to occur in real-time via the M-extractor which digitally implements Eq. (\ref{eqt:tmfunc}) as a weighted sum:

\begin{equation}\label{eqt:m-vec}
    m\text{-}\text{vector} = \sum_{i=0}^{99}\frac{w_i*h(t)}{256}
\end{equation}

\noindent where $h(t)$ is the input signal. Here, the weights $w$ are supplied by the user through a 100-element vector, each entry corresponding to an 8-bit signed coefficient scaled by 256 (thus between -0.5 and 0.5). Internally, each incoming sample is multiplied by its corresponding weight, and the resulting products are accumulated. The output is an $m$-value. 80 $m$-values make up the $m$-vector, where each $m$-value is a single 12-bit signed value representing the projection of the measured signal onto the selected mode. The M-extractor ultimately allows one to flexibly select and switch between temporal modes entirely in software, negating the requirement for any post-processing, and therefore allowing for real-time data processing.

The cost is a latency of 100 ns i.e., the M-extractor is responsible for half of the 200 ns system latency budget. It receives $4$ ADC samples every $4 \ \text{ns}$ clock cycle (cc), multiplies each by its corresponding weight (from the  $100$-element weight vector), and accumulates the results over $100$ samples. As $4$ samples arrive per cc, processing $100$ samples takes $100 \div 4 = 25$ clock cycles. With each cc lasting $4 \ \text{ns}$, the total latency is $25 \times 4 \ \text{ns} = 100 \ \text{ns}$.

\subsubsection{PM-Gain and PM-Compensation}

The DAC drives both an \gls{im} and a \gls{pm}, specifically the MX-0.1-LN intensity modulator and the MPX-0.1-LN phase modulator from Exail.
Both devices operate via the Pockels effect (linear electro-optic effect), in which an applied electric field induces a proportional change in the refractive index of a non-linear crystal. This refractive index change produces a corresponding change in the optical phase of the transmitted light.

In a phase modulator, this effect is used directly to control the phase of the light. To convert phase modulation into intensity modulation, an \gls{im} is typically implemented as a Mach–Zehnder interferometer (MZI): the optical signal is split into two arms, a phase shift is applied to one arm, and the beams are recombined. The induced phase difference between the arms leads to constructive or destructive interference, producing the desired intensity modulation. However, because an \gls{im} works by introducing a phase shift, it can also cause an unintended residual phase change in the output light when viewed in the phase space representation. As a result, when the intensity-modulated light subsequently passes through the \gls{pm} for additional phase modulation, there is an extra, unwanted phase offset originating from the \gls{im}.

The \textit{PM-Comp} setting is used to correct for this. This configurable parameter applies a compensating adjustment to cancel out the residual phase introduced by the \gls{im}, ensuring that the \gls{pm} applies only the intended phase modulation.

\begin{figure}
\centering
\includegraphics[width=0.9\linewidth]{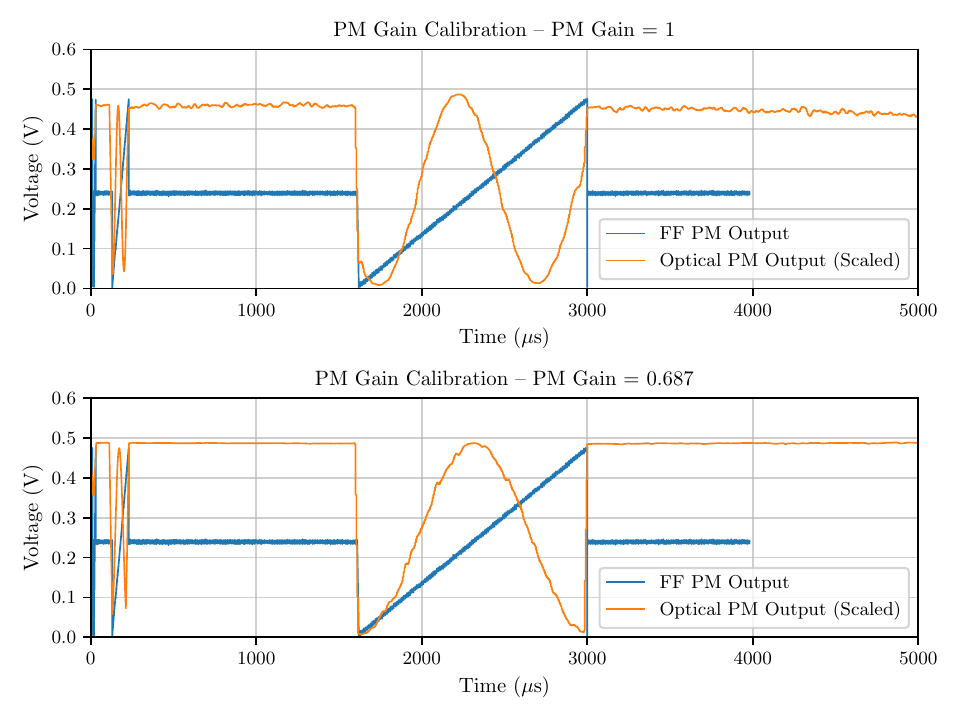}
\caption{PM-Gain Calibration. In the top figure \textit{PM-Gain} is set to 1, such that it has no affect. It is seen that for a $-\pi$ to $\pi$ (0-0.48 V) ramp output from the DAC, the \gls{pm} produces a modulation greater than 2$\pi$. With the correct value of \textit{PM-Gain} set (0.687), a 2$\pi$ modulation is correctly achieved.}
\label{fig:pm_gain}
\end{figure}

The \gls{pm} must provide phase shifts over the full range from $-\pi$ to $\pi$. For the MPX-0.1-LN, the nominal $V_{\pi}$ is 1 V, so a 2 V peak-to-peak drive is needed to span this range. Since the \gls{ff}-system DAC can output at most 0.96 V, an analog amplifier is inserted before the \gls{pm} to boost the signal. This amplification alters the effective $V_{\pi}$ as seen from the DAC’s perspective and, because both the amplifier gain and the PM’s true $V_{\pi}$ can deviate from their nominal values, the actual phase shift per DAC volt is not precisely known. The \textit{PM Gain} parameter in the PM Adjust entity compensates for this uncertainty by applying a configurable digital scaling factor to the PM drive signal, ensuring that the DAC’s 0.96 V range maps accurately to the desired $-\pi$ to $\pi$ phase range after amplification. A \textit{PM Gain} of 0.687 is required (Fig \ref{fig:pm_gain}). The \textit{PM Comp}, as it turns out, is not required for the modulators used here.

\subsubsection{Streaming of A-vectors}

The large number of $A$-vectors required for the computation cannot be stored in the \gls{fpga} BRAM due to capacity constraints. Instead, they reside in the \gls{ps} DDR4 memory, which provides a total of 4 GB of addressable space and an effective 64-bit DRAM bus width operating at 1066.56 MHz. To deliver these vectors into the \gls{fpga} fabric at the necessary rate without overloading the ARM cores, an AXI \gls{dma} engine\cite{amd_dma} is employed in Memory-Mapped to Stream (MM2S) mode.

In this configuration, the DMA fetches $A$-vector blocks from DDR via the \gls{ps_fpga}’s high-performance (HP) AXI ports and presents them as an AXI4-Stream to the fabric. The ARM cores configure the DMA by setting source addresses, transfer lengths, and start commands via its AXI4-Lite control interface, but once initiated, the transfer proceeds entirely in hardware. This offloads the processor from handling bulk data movement, freeing it to perform higher-level coordination and control tasks. The fetching is initiated by the Ax and Ap-Bram's via \textit{Read\_Burst}, with a burst consisting of 1024 vectors being sent.

The \gls{ps_fpga} memory interface and the calculation pipeline operate in different clock domains (300 MHz v 250 Mhz) necessitating a \gls{cdc}. To address this, the DMA’s AXI4-Stream output is fed into a custom designed stream\_avec entity, which implements a \gls{cdc} FIFO. This FIFO buffers incoming data in the 300 MHz domain and releases it in the 250 MHz domain using Gray-coded pointer synchronisation to safely pass write and read indices between domains. The \gls{cdc} not only provides a safe transfer but also absorbs small timing skews between the two domains, allowing sustained operation even under variable DDR access latencies.

\subsubsection{DAC Output and Voltage Mapping}

\begin{figure}
\centering
\includegraphics[width=0.9\linewidth]{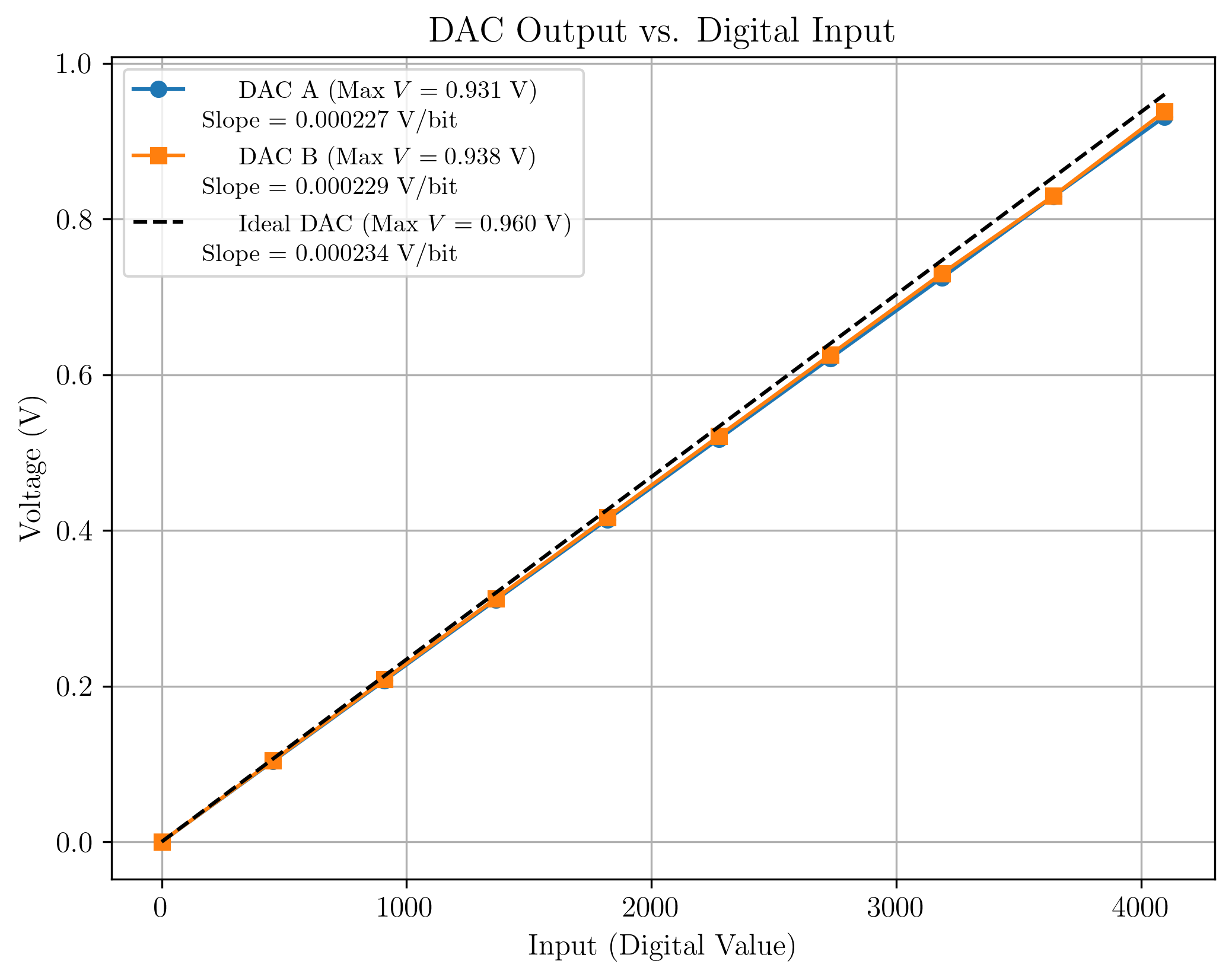}
\caption{Measurement of the maximum DAC output. 0.96 V is the theoretical maximum. The measured maximum is 0.93 for DAC A (\gls{im} output) and 0.94 for DAC B (\gls{pm} output)}
\label{fig:dac_output}
\end{figure}

The system contains two DAC outputs, one to drive the \gls{im} (DAC A) and the other to drive the \gls{pm} (DAC B). The measured DAC output range is 0 V to 0.93 V for DAC A and 0.94 V for DAC B (Fig \ref{fig:dac_output}). The maximum voltage of 0.93 V is mapped to a value of 2 (0xFFF) for the \gls{im} output, where the output is in fixed point representation. However the CORDIC is unable to produce a value larger than $\sqrt{2}$ for the magnitude. Thus, the maximum voltage DAC A will output is 0.679 V, which fits inside 1.11.

\begin{figure}
\centering
\includegraphics[width=0.9\linewidth]{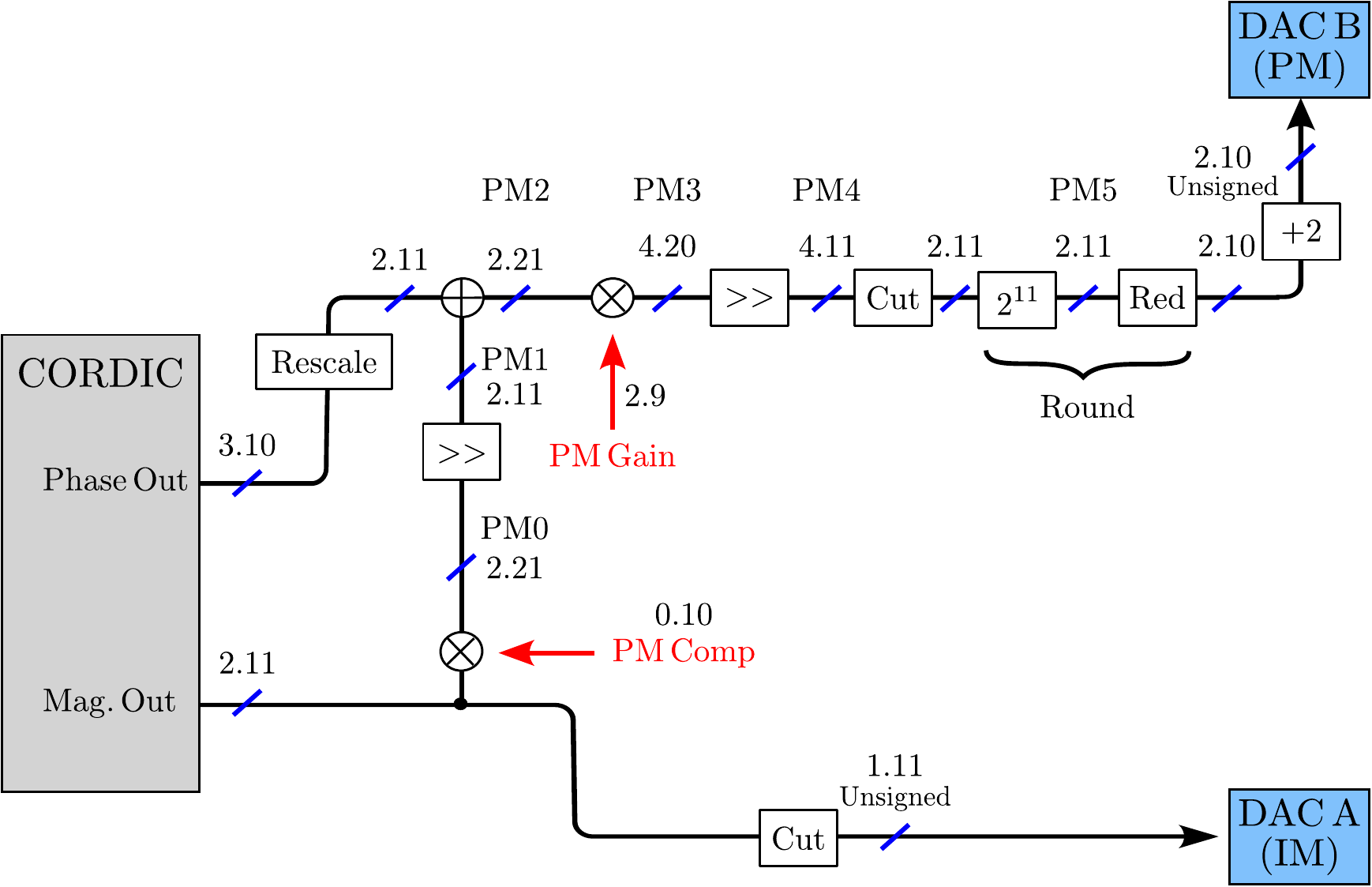}
\caption{Schematic of the PM Adjust entity. The CORDIC outputs a phase in the range -1 to +1 in 3.10 format, which is rescaled to 2.11 format. The magnitude value is output in 2.11 format. To compensate for phase changes due to the \gls{im}, a user input \textit{PM Comp} is available which ranges from -0.5 to +0.5. This value is multiplied by the CORDIC magnitude output (1.141 $\approx$ 1.5), producing a maximum value of 0.707 ($\approx$ 0.75), stored in the internal signal PM0. The result is added to the CORDIC phase output, yielding a maximum range of -1.75 to +1.75. The \textit{PM Gain} parameter (range: -1 to +1) is then multiplied with this signal. Afterward, the data undergoes bit manipulation and rounding, before being offset by 1024 to render the output unsigned.}

\label{fig:PM_adjust}
\end{figure}

For the phase output (DAC B), the numeric mapping is defined to accommodate the phase-modulation compensation i.e. PM Adjust, which actually happens inside the PM Adjust entity itself (Fig. \ref{fig:PM_adjust}). Without PM Adjust, a numeric range of -1.000 to +1.000 from the CORDIC phase output corresponds to a physical phase range of $-\pi$ to $+\pi$, i.e., a total span of $2\pi$. With the addition of the PM Adjust block, the phase value is first corrected for amplitude-phase coupling and then scaled by the user-adjustable \textit{PM Gain} parameter. This correction can shift the phase by up to approximately $\pm 0.75\pi$ in addition to the nominal phase. To prevent numerical saturation in the DAC input when both terms are present at their maximum, the internal mapping is redefined such that a numeric range of -2.000 to +2.000 corresponds to $-2\pi$ to $+2\pi$, i.e., a total phase span of $4\pi$. 

This doubling of the numeric span provides sufficient headroom for the compensated phase (-1.75 to +1.75) without clipping, at the expense of one bit of effective resolution in the DAC input code. Further, as the DAC cannot output negative values, it is offset by +1024. The measured maximum output voltage for DAC B is 0.94 V (Fig. \ref{fig:dac_output}), which corresponds to the numeric value +2.000 in the mapping ($2\pi$). Therefore $+\pi$ corresponds to 0.470 V, 0 to 0.235 V and $-\pi$ to 0 V, due to the offset. Consequently, when the \textit{PM\_Gain} parameter is set to its default value of $+1.000$, the phase output covers the extended $4\pi$ range, ensuring the full dynamic range of the compensated phase is preserved in the DAC output. Again, however, as the CORDIC phase output is always between $-\pi$ and $\pi$, the maximum voltage the DAC outputs is 0.47 V.

\begin{table}[h]
\centering
\caption{DAC output voltage mapping for IM and PM channels. Theoretical values assume a full-scale output of 0.96 V.}
\begin{tabular}{c c c}
\hline
\textbf{Magnitude} & \textbf{IM Voltage (theoretical)} & \textbf{IM Voltage (measured)} \\
\hline
0         & 0 mV   & 0 mV   \\
1.0       & 480 mV & 467 mV \\
$\sqrt{2}$ & 679 mV & 660 mV \\
\hline
\end{tabular}
\quad
\begin{tabular}{c c c}
\hline
\textbf{Phase} & \textbf{PM Voltage (theoretical)} & \textbf{PM Voltage (measured)} \\
\hline
$-180^\circ$ & 0 mV   & 0 mV   \\
$0^\circ$    & 240 mV & 235 mV \\
$+180^\circ$ & 480 mV & 470 mV \\
\hline
\end{tabular}
\label{tab:dac_mapping}
\end{table}

\section{Verification}\label{sec:verification}

The system is verified in simulation and hardware. Individual entities were verified at time of design, with a full system testbench also developed. All testbenches were designed using System Verilog. A second verification test using pulsed light and homodyne detection is described in appendix \ref{sec:veri_pulse}.

\begin{figure*}[!ht]
 \centering
  {\includegraphics[width=\linewidth]{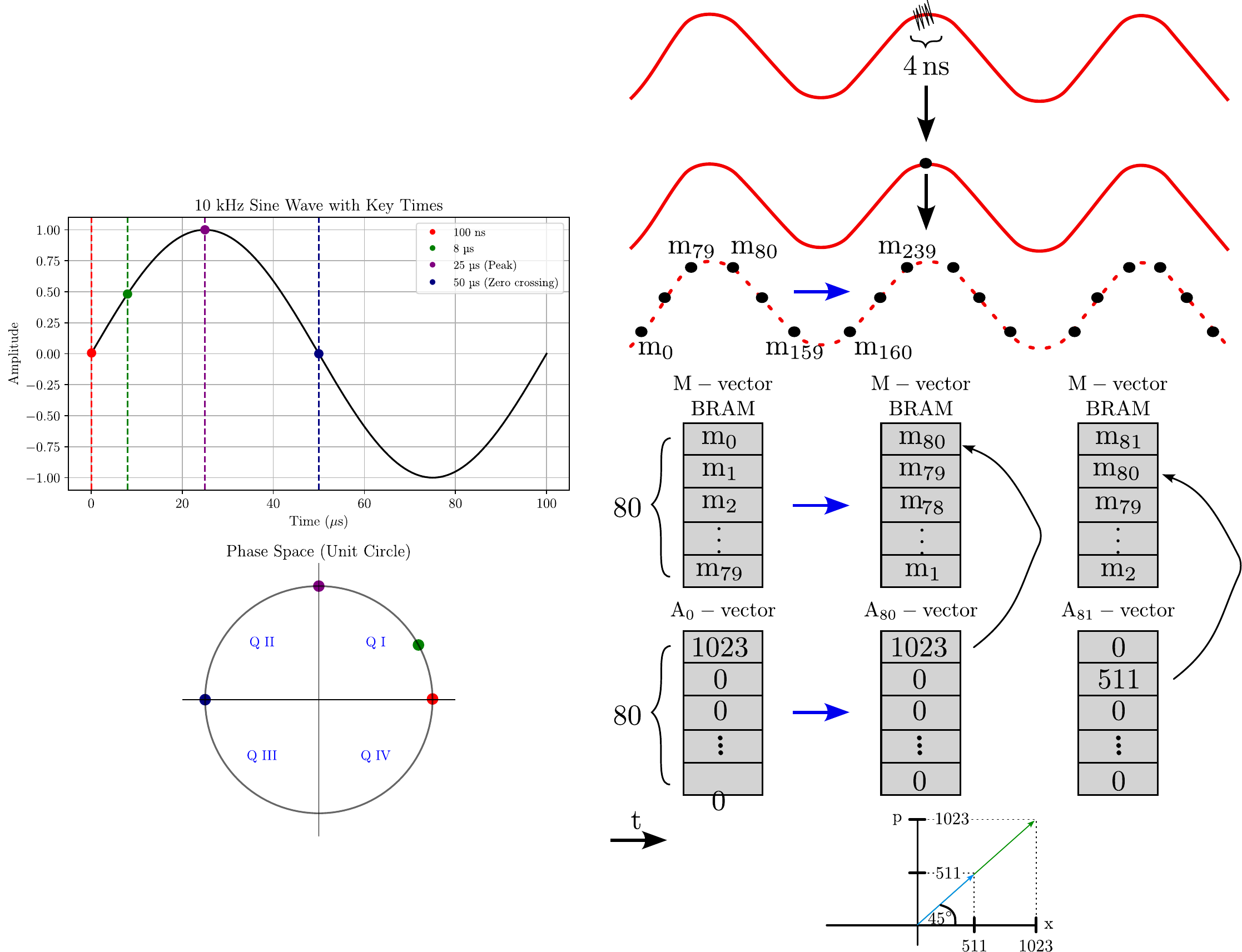}\label{fig:calc_veri_pic}}
   \caption{Understanding the system calibration test. A 10 kHz sine wave is input, with an $m$-value created every 100 ns and fed into the $M$-vector shift register with a gain of 1. The same positive $m$-value is multiplied by 80 $A_x$ vectors and 80 $A_p$ vectors, alternating between 1023 and 511, and takes $8,\mu\mathrm{s}$ to pass through the shift register. In phase space (the unit circle), the operating point is therefore in Quadrant I, as both inputs are positive. The phase output of the CORDIC is consequently a constant $45^\circ$. After $50,\mu\mathrm{s}$, the $m$-values become negative as the sinusoid enters its negative half-cycle. The operating point therefore moves to Quadrant III in phase space, and the CORDIC phase output becomes $-135^\circ$. As such, the phase output should oscillate between $45^\circ$ ($\approx 290$ mV) and $-135^\circ$ ($\approx 59$ mV) every $50,\mu\mathrm{s}$ if the system is operating correctly. Incorrect fetching of $A$ vectors or misaligned data would appear as significant deviations in the phase output.}

\label{fig:calc_expl}
\end{figure*}

The hardware and full-system verification consisted of inputting a 10 KHz, 2 Vpp sinus into the ADC; via a signal generated in System Verilog in the simulation and an Arbitrary Waveform Generator in the hardware test. The M-extractor weights were set to [64, 64, 64, 64, 0, 0, ..., 0], i.e., set to subsample the ADC input down to 10 MHz with gain of 1 (64$\cdot$4 = 256). The $A$-vectors consisted of a repeating pattern; the first $A$-vector had a value of 1023 at the last index ($A_{79}^0$), the second vector a value of 511 at the second last index ($A_{78}^1$), the third vector a value of 1023 at the third last index ($A_{77}^2$) etc. . As the $m$-vector is a shift register, each $A$-vector therefore multiplied the same $m$-value (Fig. \ref{fig:calc_expl}) for 8 µs. 80,100 $A_x$-vectors and 80,100 $A_p$-vectors were written to the DDR to support a 8.000 ms long measure period. This pattern was duplicated for both $A_x$-vectors and $A_p$-vectors.

\begin{figure*}[!t]
    \centering
    \subfloat[]{\includegraphics[width=0.9\textwidth]{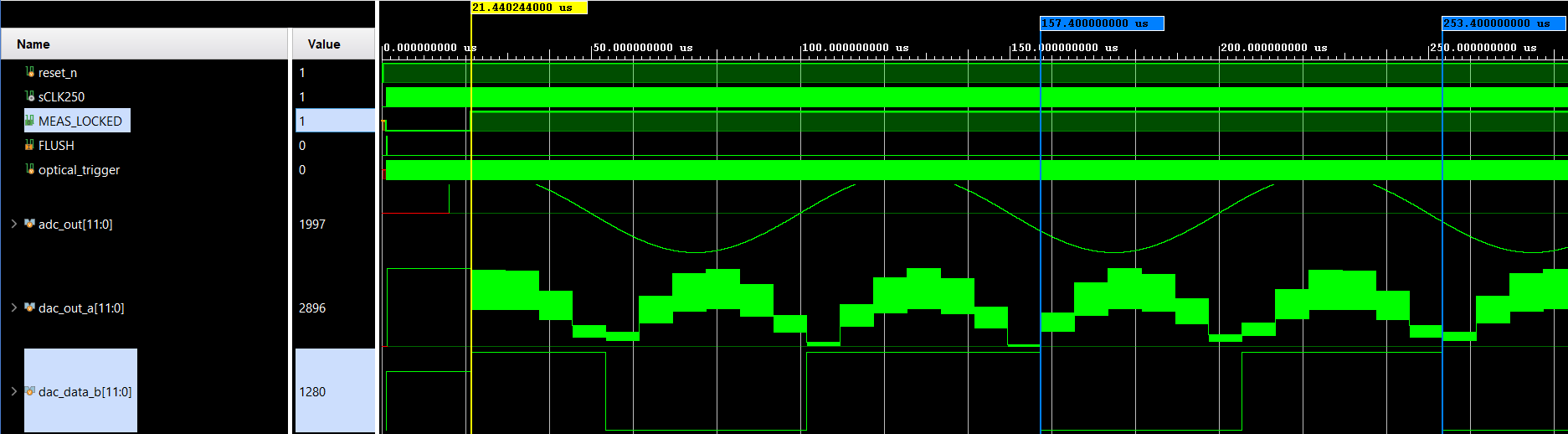}\label{fig:calc_veri_vivado}}\\
    \subfloat[]{\includegraphics[width=0.9\textwidth]{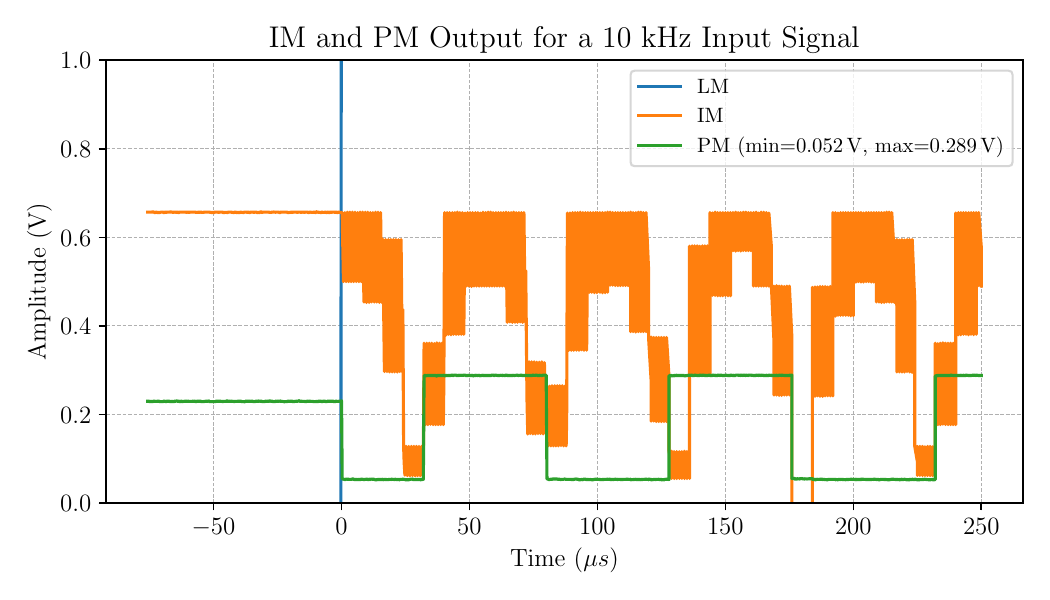}\label{fig:calc_veri_osci}}
    \caption{Vivado simulation (top) and experimental data (bottom) of the system verification test. It is seen that the phase output correctly oscillates every $50,\mu\mathrm{s}$ from $45^\circ$ to $-135^\circ$. A zoomed-out version is provided in Appendix~\ref{sec:calc_val_appendix}. LM = lock-measure signal, IM = DAC A output, and PM = DAC B output.}

   \label{fig:calc_veri_results}
\end{figure*}

This setup was designed to cause the phase output to oscillate between $45^\circ$ and $-135^\circ$. As the $A$-vector setup was duplicated on the $x$ and $p$ side of the equation, the same absolute values of $x$ and $p$ entered the CORDIC. Therefore, the phase output $\alpha=\arctan\left(\frac{p}{x}\right)$ was always equal to $45^\circ$ when the sinusoid was positive (thus positive $x$ and $p$ values), and $-135^\circ$ when the inputs were negative (negative $x$ and $p$ values). This is due to the rectangular-to-polar coordinate transformation. When both inputs are positive, the point lies in Quadrant I and the phase ($\theta$) is equal to $\alpha$. When both inputs are negative, the point lies in Quadrant III and the phase ($\theta$) is equal to $-(180^\circ-\alpha)=-(180^\circ-45^\circ)=-135^\circ$. As the 10 kHz sinusoid has a period of $100,\mu\mathrm{s}$, it becomes negative every $50 \, \mathrm{\mu s}$. Therefore, the phase should oscillate between $45^\circ$ and $-135^\circ$ every $50\, \mathrm{\mu s}$. Fig.~\ref{fig:calc_veri_results} shows the measured data alongside the Vivado simulation. The phase output is as expected.

The shifted 1023/511 $A$-values cause 100 ns \gls{im} output oscillations between 1/2 and 1/4 of the $m$-values value. This occurs for 8 µs until this $m$-value exits the shift register. A new $m$-value enters and the process repeats. The \gls{im} output follows the sinusoid shape. 

The test ensures that any missed $A$-vector, data misalignment or calculation issue would give a substantially different phase output or magnitude output (see appendix \ref{sec:calc_valid_avec}). Using 1023 instead of 1024 stressed the multiplier. As 80 does not divide the 4096 BRAM buffer size, a DMA overrun or under-run would be detected by A-vectors being shifted in time or missing altogether giving incorrect \gls{im} output. 

\begin{figure}[!t]
\centering
\includegraphics[width=0.8\columnwidth]{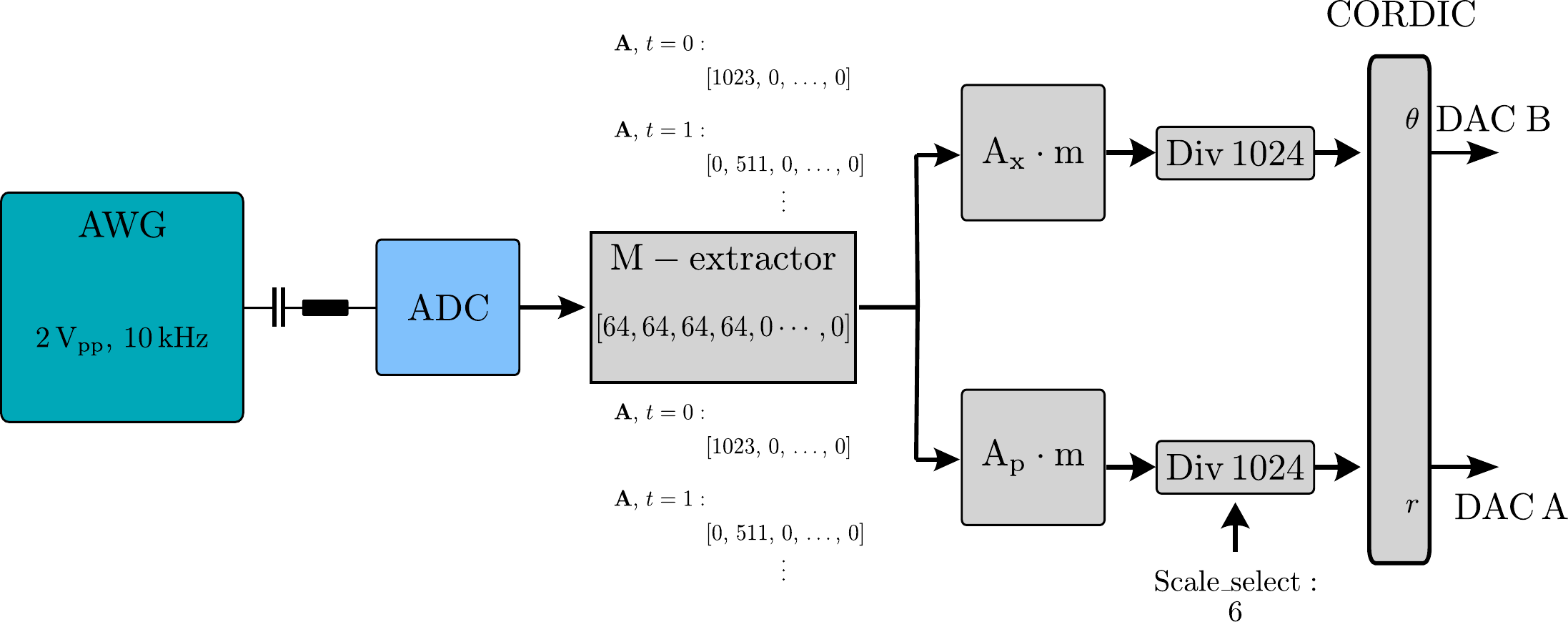}
\caption{Schematic of the \gls{ff}-system verification set-up.}
\label{fig:ff_system_setup}
\end{figure}

\begin{figure*}[!t]
    \centering
    \subfloat[]{\includegraphics[width=0.9\textwidth]{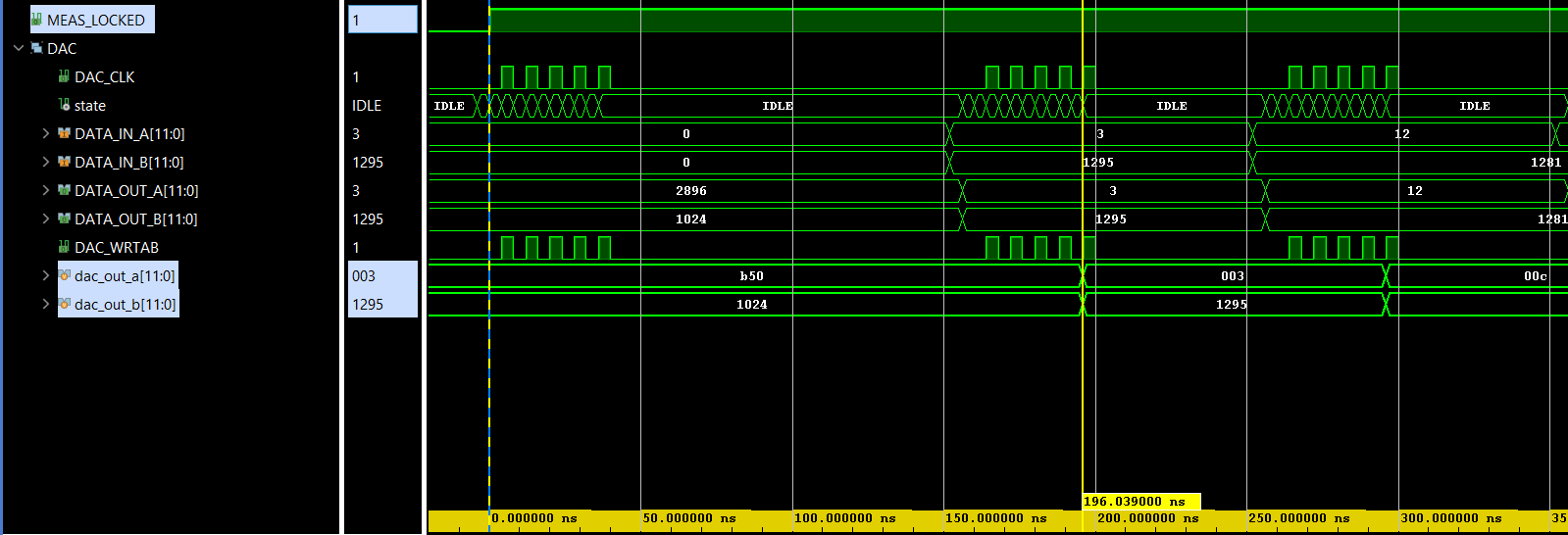}\label{ls}}\\
    \subfloat[]{\includegraphics[width=0.6\textwidth]{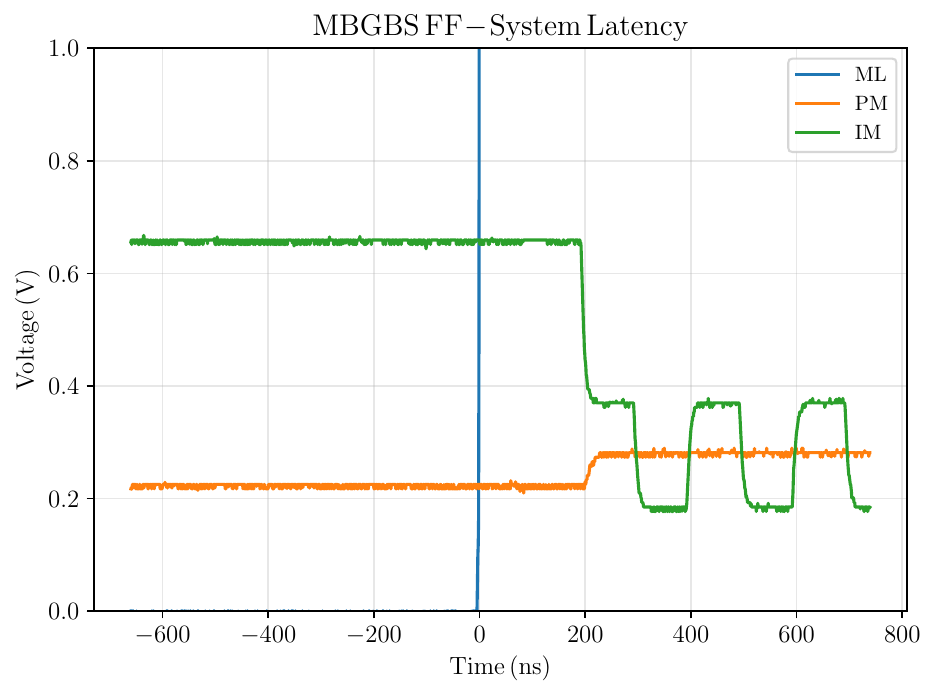}\label{fig:lat}}
    \caption{Vivado simulation (top) and experimental data (bottom) of the system latency, measured to be 196 ns.}
   \label{fig:latency}
\end{figure*}

\section{Discussion}

The \gls{fpga}-based fast \gls{ff} system presented here demonstrates that real-time processing of homodyne measurement outcomes is feasible with latencies compatible with \gls{cv} \gls{mbqip} and photonic simulation protocols. With respect to \gls{cv} \gls{mbqip}, post-processing approaches limit scalability and propose practical limitations. Our implementation shows that FPGA hardware can bridge the gap between theory and experiment by enabling deterministic, low-latency corrections essential for continuous-variable quantum computation to occur in real time, and allow for the required scalability.

A central outcome of this work is the demonstration of flexible and programmable inner-product operations between measurement vectors and pre-stored matrices as well as precise timing and experimental control. These capabilities directly support photonic quantum computing. The system provides the low-latency displacements required to undo conditional shifts induced by homodyne measurements. Our system achieves a total latency of 196 ns, with half of this attributed to the M-extractor module, operating on a 250 MHz clock. This latency is compatible with delay lines of approximately 40 meters per photon assuming a bulk optical setup and standard 1550 nm optical fiber. 

Typically one wants to translate bulk-optical experiments to integrated experiments as it allows for greater scalability. However, integrated implementations pose greater challenges. Propagation losses are far larger (1–2 dB/cm), making photon delay management more difficult. In such scenarios, electronic latency becomes far more constrained, as achieving delays longer than 2 ns is experimentally challenging. Further, higher repetition rate lasers are needed. One potential way to overcome this is for the delay line to exist off chip, allowing for longer delays. Our system contains all the necessary components to work with chip-based implementations, but requires at least a doubling of the clock speed counting for the fact that the laser repetition rate will be increased. Further, some entities, such as the multiplier, may require pipelining as a result.

Finally, the system's underlying architecture is general-purpose. The same principles could be extended to other measurement-driven quantum protocols, adaptive optics, or real-time Hamiltonian engineering tasks, such as that required in topological photonics. This highlights the broader relevance of FPGA-based feedforward beyond the immediate applications suggested here. Future efforts will focus on optimising the system for integrated platforms and expanding its capabilities to support more complex quantum protocols and simulations.

\section{Conclusion}

We have provided a \gls{fpga}-based fast feedforward system designed to meet the stringent latency requirements of \gls{cv} \gls{mbqip} and topological photonics applications. The system integrates a \gls{hqe} homodyne detector, low-latency signal processing, and real-time feedforward capabilities, achieving a total system latency of less than 200 ns. By leveraging the parallel architecture and reconfigurability of FPGAs, the system performs critical inner-product calculations and phase-space displacements essential for scalable quantum computing protocols.

Key contributions include the development of a high-performance fully fiber-based homodyne detector with 15 dB clearance at 1 GHz and quantum efficiencies exceeding 95\%, as well as a robust FPGA architecture capable of processing measurement outcomes and generating control signals within sub-microsecond timescales.

\section*{Acknowledgment}
The authors would like to thank Zhenghao Liu, Olga Solodovnikova, Romain Brunel and Oscar Cordero Boronat for valuable discussions, both technical and physical. This work is supported by CLUSTEC from the EU Horizon Europe program No. 101080173 and EPIQUE from the EU Horizon Europe program No. 101135288.

\clearpage

\appendices

\section{Mathematical Description of Feedforward System Calculation}\label{sec:calck_math}

\begin{figure}[H]
\centering
\includegraphics[width=0.85\columnwidth]{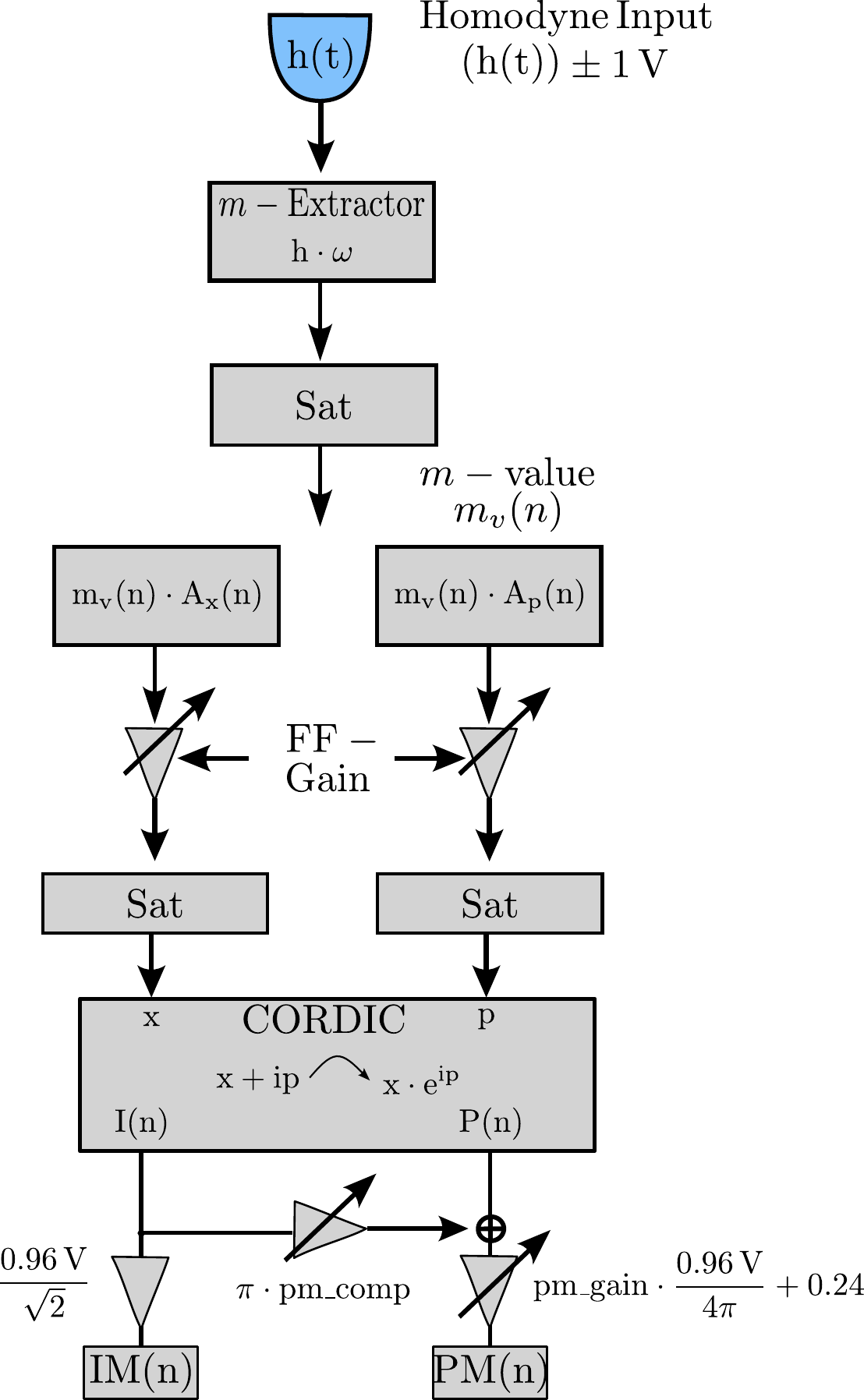}
\caption{Schematic of the \gls{ff}-system calculation as mathematical equations.}
\label{fig:ff_calc_math}
\end{figure}

Below we give the mathematical calculation that the \gls{ff}-system undertakes (see Fig. \ref{fig:ff_calc_math}).

\begin{equation}
\begin{aligned}
&\text{Sat}(x) = 
\begin{cases}
+1, & x > 1 \\
x, & -1 \leq x \leq 1 \\
-1, & x < -1
\end{cases} \\
&w_k \in [-0.5, 0.5), \quad h(t) \in [-1\,\text{V}, 1\,\text{V}] \\
&A_{x_{n,k}}, A_{p_{n,k}} \in [-1, 1]
\end{aligned}
\end{equation}

\begin{equation}
m_v(n) = \text{Sat}\left\{ \sum_{k=0}^{99} w_k \cdot h(n \cdot 100\,\text{ns} + k \cdot 1\,\text{ns}) \right\}
\end{equation}

\begin{equation}
x(n) = \text{Sat} \left\{ \text{ff-gain} \cdot \sum_{k=0}^{79} m_v(n - 79 + k) \cdot A_{x_{n,k}} \right\}
\end{equation}

\begin{equation}
p(n) = \text{Sat} \left\{ \text{ff-gain} \cdot \sum_{k=0}^{79} m_v(n - 79 + k) \cdot A_{p_{n,k}} \right\}
\end{equation}

\begin{equation}
I(n) e^{i P(n)} = x(n) + i \, p(n)
\end{equation}

\begin{equation}
I(n) \in [0, \sqrt{2}], \quad P(n) \in [-\pi, \pi)
\end{equation}

\begin{equation}
\text{pm.gain} \in [-2, 2], \quad \text{pm.comp} \in [-0.5, 0.5]
\end{equation}

\begin{equation}
\text{IM}(n) = \frac{0.93\,\text{V}}{2} \cdot I(n)
\end{equation}

\begin{equation}\label{eqt:pm_map}
\text{PM}(n) = \frac{0.94\,\text{V}}{4\pi} \cdot \text{pm.gain} \cdot \left( P(n) + \pi \cdot \text{pm.comp} \cdot I(n) \right) + 0.24\,\text{V}
\end{equation}

\section{Calculation Validation - 'Incorrect' Reading of A-vector}\label{sec:calc_valid_avec}

\begin{figure}[H]
\centering
\includegraphics[width=0.9\linewidth]{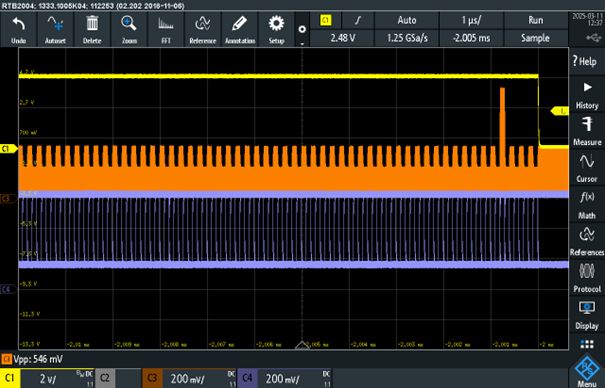}
\caption{Screenshot of the stability test after 2 hours.}
\label{fig:system_stab}
\end{figure}

The system verification test was let run for 2 hours to check stability. For this test, vector 79990 was given a value of 1500 at index 0. The large peak in Fig. \ref{fig:system_stab} represents this phase change. Such a peak would result in Fig. \ref{fig:calc_veri_results} in the system verification test (section \ref{sec:verification}) if $A$-vectors were incorrectly fetched or misaligned.

\section{Calculation Verification Experimental Data}\label{sec:calc_val_appendix}

Below is a 'zoomed out' version of the bottom graph in Fig. \ref{fig:calc_veri_results}.

\begin{figure}[H]
\centering
\includegraphics[width=0.9\linewidth]{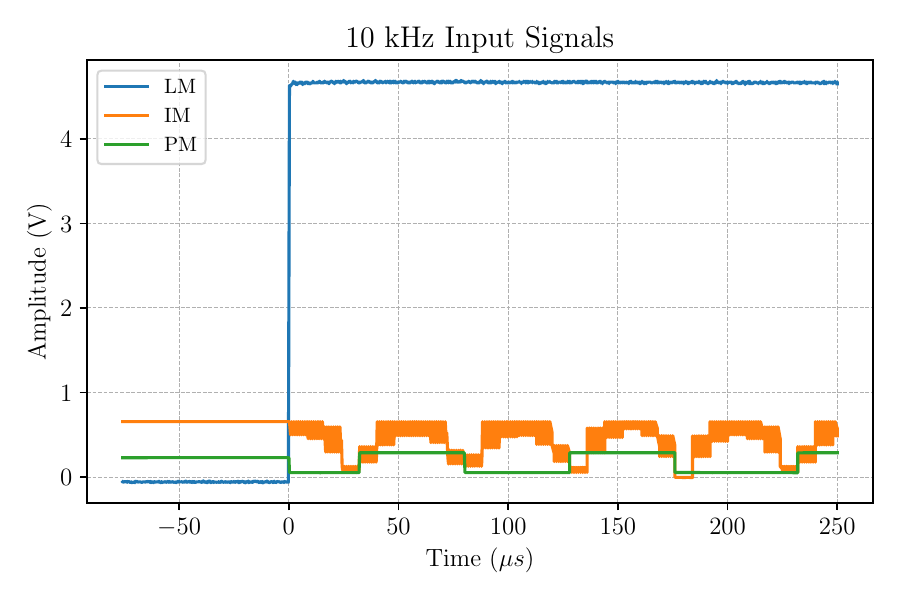}
\caption{Calculation validation data over 250 µs.}
\label{fig:calc_val_results_far}
\end{figure}

\section{Timing Summary and Resource Utilisation}\label{sec:resource util}

Below we show the timing summary and the resource utilisation from the implemented design. All data is taken from Vivado.

\begin{table}[ht]
\centering
\caption{Design Timing Summary}
\begin{tabular}{lccc}
\hline
\textbf{Metric} & \textbf{Setup} & \textbf{Hold} & \textbf{Pulse Width} \\
\hline
Worst Slack (ns) & 0.020 & 0.009 & 0.100 \\
Total Slack (ns) & 0.000 & 0.000 & 0.000 \\
Failing Endpoints & 0 & 0 & 0 \\
Total Endpoints & 162716 & 162716 & 44091 \\
\hline
\end{tabular}
\label{tab:timing_summary}
\end{table}

\begin{figure}[H]
\centering
\includegraphics[width=0.9\linewidth]{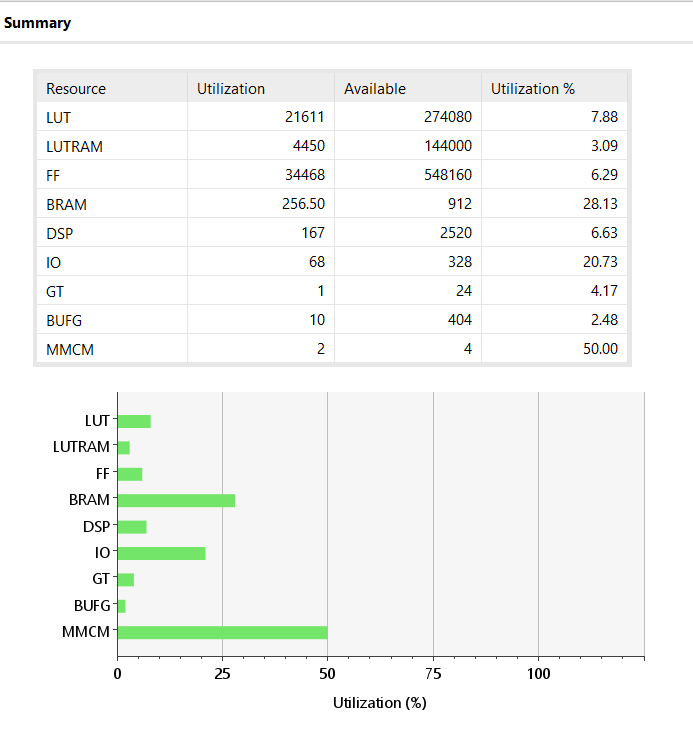}
\caption{Summary of resources used by the implemented design. Screenshot from Vivado.}
\label{fig:resource utilisation.}
\end{figure}

\section{Writing from DDR to Fabric BRAM}\label{sec:write_avecs}

The writing of data to DDR and then to the BRAM containing the A-vectors is confirmed to work. The input data to the DDR is a repeating pattern of: 
'FFF', '001', 'FFE', '002', 'FFD', '003', 'FFC', '004', 'FFB', '005', 'FFA', '006', 'FF9', '007', 'FF8', '008'.

\begin{figure}[!htbp]
\centering
\includegraphics[width=\linewidth]{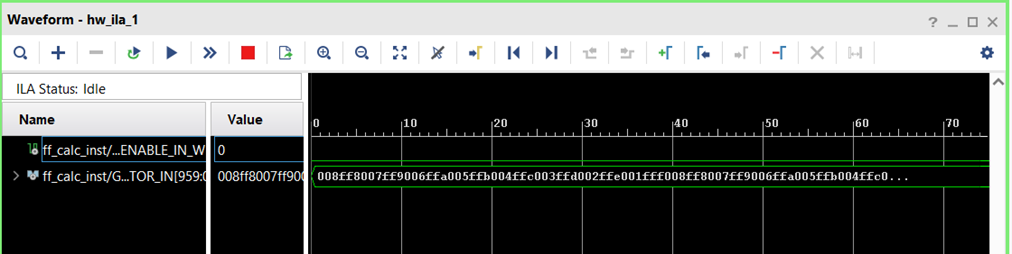}
\caption{Screen shot of data inside the Ax-BRAM. Screenshot from Vivado. Data taken using an Integrated Logic Analyser (ILA). }
\label{fig:data_bram}
\end{figure}

The data inside the DDR is shown in Fig. \ref{fig:data_ddr}. The resultant input to the BRAM is shown to be correct in Fig. \ref{fig:data_bram}.

\begin{figure}[!htbp]
\centering
\includegraphics[width=0.9\linewidth]{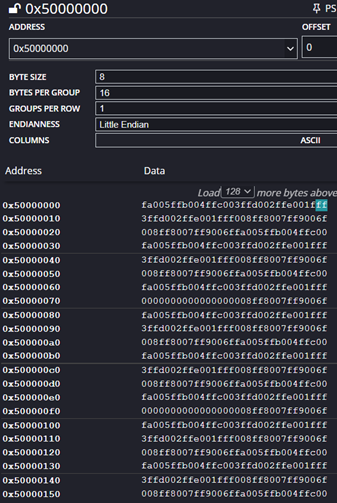}
\caption{Screen shot of Vitis' memory inspector output showing the data inside the DDR. The first byte is highlighted. }
\label{fig:data_ddr}
\end{figure}

\section{Verification using Pulsed Light}\label{sec:veri_pulse}

The feedforward system is verified using a 10 MHz pulses.

The DAC is 12 bit, so can output a max digital value of 2$^{12}$ = 4096. DAC A is the \gls{im}-output. DAC B is the \gls{pm}-output. The max voltage DAC A can output is 0.93 V. So 0.93 V = 4096. For DAC B; 0.94 V=4096.

\begin{figure}[!ht]
\centering
\includegraphics[width=0.9\linewidth]{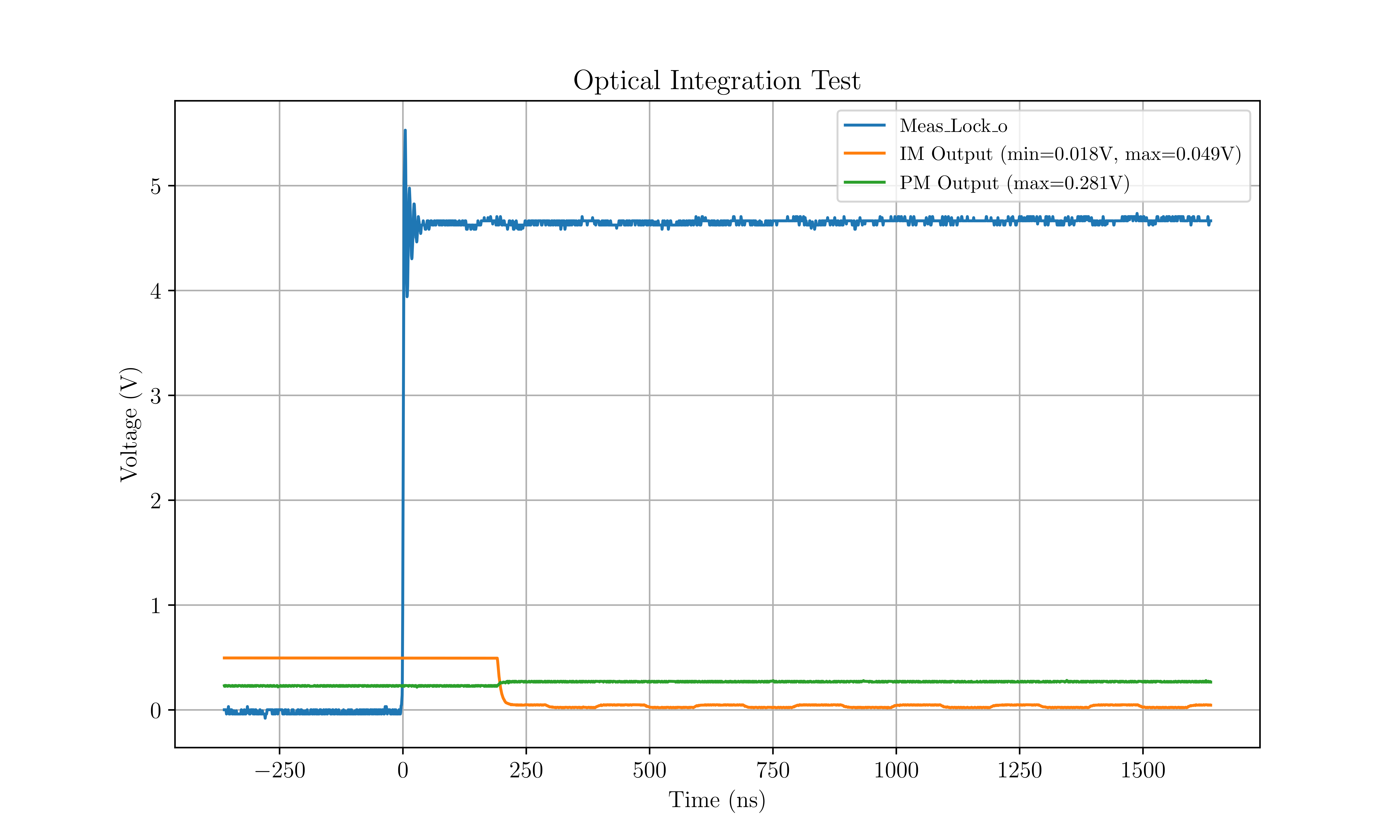}
\caption{The system outputs the correct voltages verifying it works as expected. Output occurs 200 ns after the Lock\_meas\_o signal is high, due to the 200 ns system latency.}
\label{fig:opt:int_test_output}
\end{figure}

In simulation, DAC A output is a digital value of 184 and 92 i.e. (184/4096) $\cdot$ 0.93 V = 0.042 V and (82/4096) $\cdot$ 0.93V = 0.021 V. DAC B output is a digital value of 1280 i.e.  (1280/4096) $\cdot$ 0.93V = 0.29 V. Therefore the expected output in hardware is: DAC A = 0.042 V and 0.021 V ; DAC B = 0.29 V. It is shown in Fig. \ref{fig:opt:int_test_output} that the required voltages are output indicating the system works as required.

\begin{figure}[!htbp]
\centering
\includegraphics[width=0.9\linewidth]{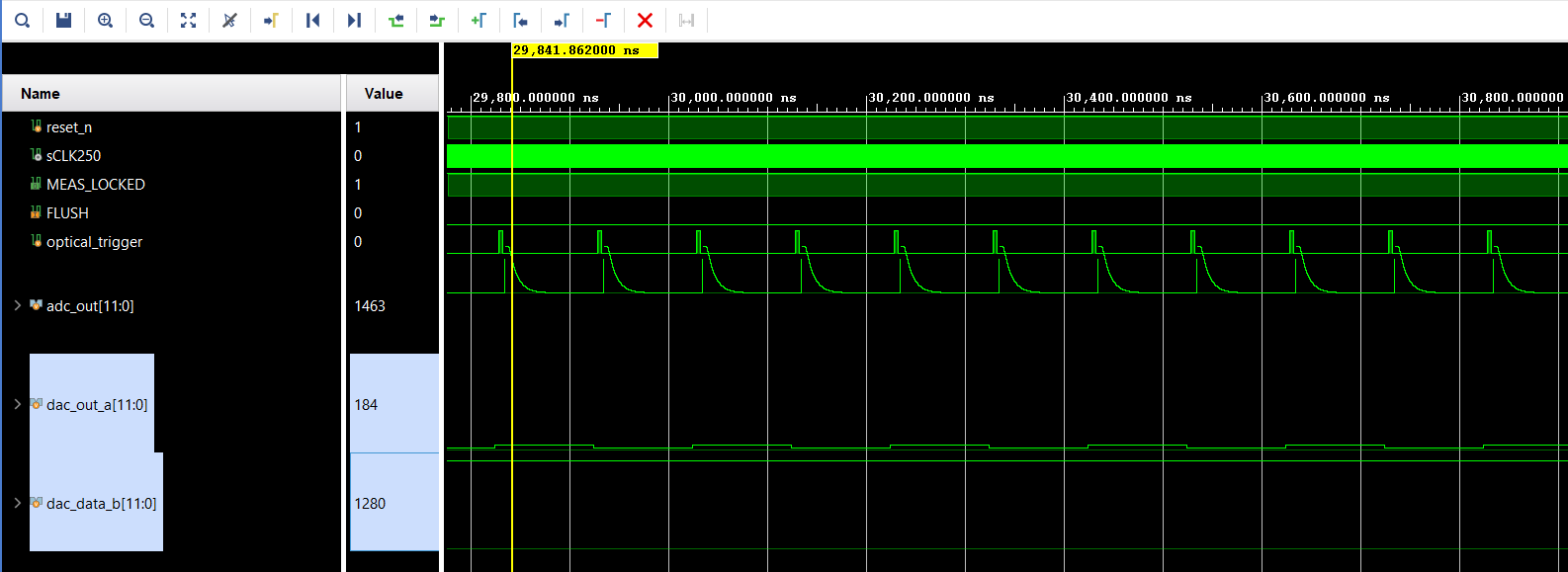}
\caption{Screenshot from the Vivado simulation of the expected output when the input is a 10 MHz train of pulses. The DAC digital values are shown. The input signals are made to represent the signals seen in Fig. \ref{fig:opt_input_hd}.}
\label{fig:data_ddr}
\end{figure}

\begin{figure}[!htbp]
\centering
\includegraphics[width=0.9\linewidth]{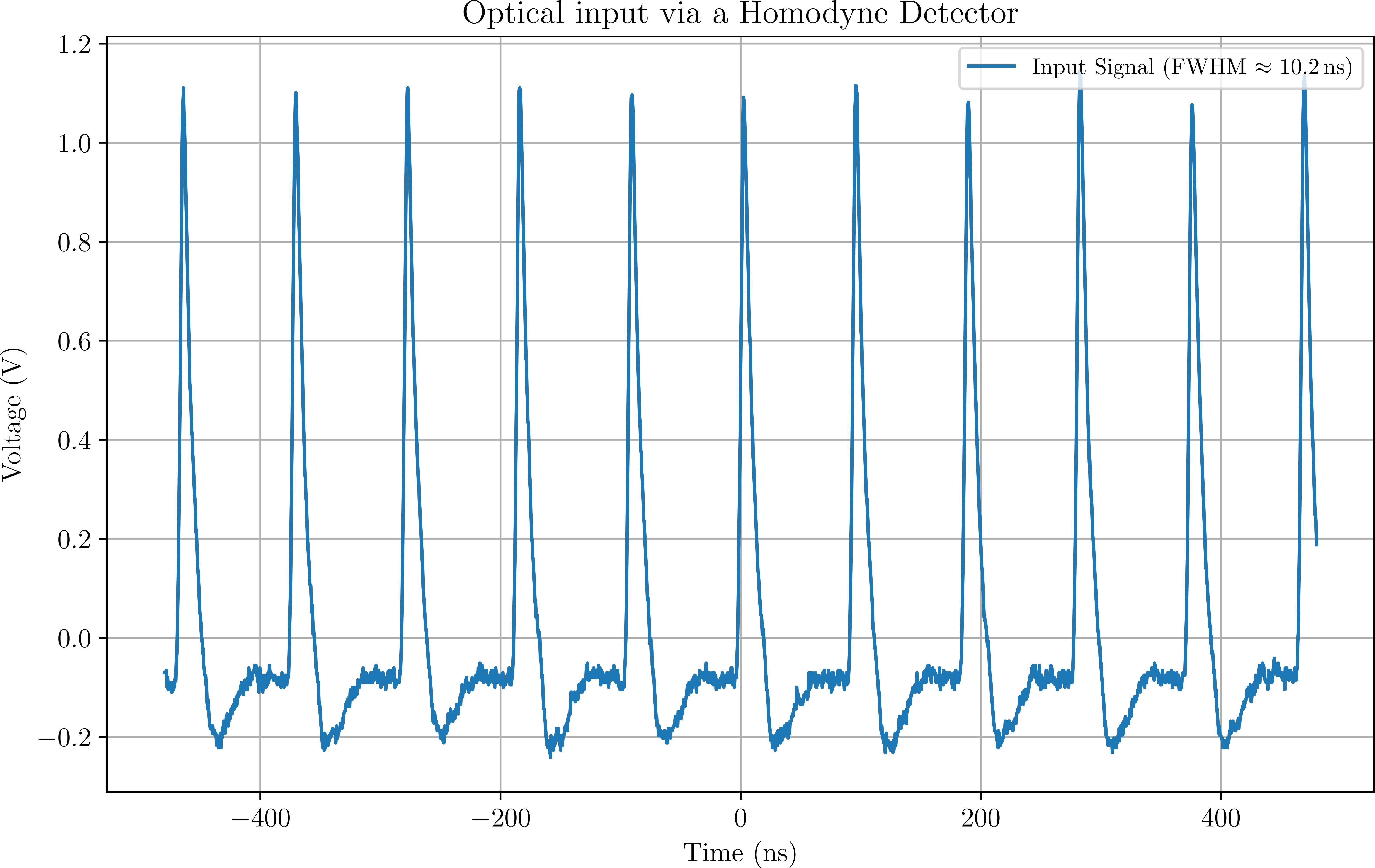}
\caption{The optical signals input to the \gls{ff}-system via a \gls{hd} are shown. These signals have a FWHM of approximately 10 ns. }
\label{fig:opt_input_hd}
\end{figure}

The \gls{ff}-system is seen to output the expected voltages value, confirming it operates as required.
\clearpage

\bibliography{Bibliography/bib}

\end{document}